%% file: ps_DMFT_1109_5_submit.tex
\documentclass[aps,prb,twocolumn,superscriptaddress,a4paper]{revtex4-1}
\usepackage[dvipdfmx]{graphicx}

\usepackage{amsmath,amsthm,amssymb}
\usepackage{amsmath,bm}
\usepackage{color}
\usepackage{ifthen}

\usepackage{multirow,bigdelim}

\newcount \commentout
\newcommand{\1}{\mbox{1}\hspace{-0.25em}\mbox{l}}

\newcommand{\twonorm}[1]{|\!|{#1}|\!|_2}

\newlength{\figwidth}
\newlength{\figlarge}
\begin{document}
\title{
R-DMFT study of a non-Hermitian skin effect for correlated systems: analysis based on a pseudo-spectrum
}

\author{Tsuneya Yoshida}
\affiliation{Department of Physics, University of Tsukuba, Ibaraki 305-8571, Japan}
\date{\today}
\begin{abstract}
We analyze a correlated system in equilibrium with special emphasis on non-Hermitian topology inducing a skin effect. 
The pseudo-spectrum, computed by the real-space dynamical mean-field theory, elucidates that additional pseudo-eigenstates emerge for the open boundary condition in contrast to the dependence of the density of states on the boundary condition.
We further discuss how the line-gap topology, another type of non-Hermitian topology, affects the pseudo-spectrum. 
Our numerical simulation clarifies that while the damping of the quasi-particles induces the non-trivial point-gap topology, it destroys the non-trivial line-gap topology. 
The above two effects are also reflected in the temperature dependence of the local pseudo-spectral weight.
\end{abstract}
\maketitle

\section{Introduction}
\label{sec: intro}
In these decades, topological properties of condensed matters enhance their significance~\cite{Kane_Z2TI_PRL05_1,Kane_Z2TI_PRL05_2,HgTe_Bernevig06,Konig_QSHE2007,Qi_TQFTofTI_PRB08,TI_review_Hasan10,TI_review_Qi10}. 
Among a variety of topological systems, recently, non-Hermitian systems attract broad interest because of their unique topological phenomena induced by non-Hermiticity~\cite{Hatano_PRL96,Hu_nH_PRB11,Esaki_nH_PRB11,TELeePRL16_Half_quantized,KFlore_nHSkin_PRL18,EElizabet_PRBnHSkinHOTI_PRB19,Gong_class_PRX18,Kawabata_gapped_PRX19,Zhou_gapped_class_PRB19,Yokomizo_BBC_PRL19,Okuma_BECpg_PRL19,Yokomizo_NBlochBBCEP_arXiv20,Bergholtz_Review19,Yoshida_nHReview_PTEP20,Ashida_nHReview_arXiv20}.
Representative examples of non-Hermitian topological phenomena are the emergence of exceptional points~\cite{HShen2017_non-Hermi,YXuPRL17_exceptional_ring,Budich_SPERs_PRB19,Okugawa_SPERs_PRB19,Yoshida_SPERs_PRB19,Zhou_SPERs_Optica19,Kawabata_gapless_PRL19} and a skin effect~\cite{TELeePRL16_Half_quantized,Alvarez_nHSkin_PRB18,SYao_nHSkin-1D_PRL18,SYao_nHSkin-2D_PRL18,Lee_Skin19,Zhang_BECskin19,Okuma_BECskin19,Okuma_PS_PRB20,Mu_MbdySkin_PRB20,Lee_MbdySkin_arXiv20,Okugawa_HOSkin_arXiv2020,Kawabata_HOSkin_arXiv2020,Fu_HOSkin_arXiv2020}. 
The former is non-Hermitian topological band touching points in the bulk on which the Hamiltonian becomes non-diagonalizable. 
The latter results in extreme sensitivity to the boundary conditions.
The platform of non-Hermitian topology extends to a wide range of systems with gain and loss, e.g., photonic crystals~\cite{Ruter_nHExp_NatPhys10,Regensburger_nHExp_Nat12,Zhen_AcciEP_Nat15,Hassan_EP_PRL17,Zhou_FermiArcPH_Science18,Takata_pSSH_PRL18,Ozawa_TopoPhoto_RMP19,Xiao_nHSkin_Exp_NatPhys19}, open quantum systems~\cite{Diehl_DissCher_NatPhys11,Bardyn_DissCher_NJP2013,Rivas_DissCher_PRB13,Budich_DissCher_PRA15,Budich_DissCher_PRB15,TELeePRL16_Half_quantized,YXuPRL17_exceptional_ring,ZPGong_PRL17,Lieu_Liouclass_PRL20,Yoshida_nHFQH19,Yoshida_nHFQHJ_PRR20}, mechanical meta-materials~\cite{Yoshida_SPERs_mech19,Ghatak_Mech_nHskin_arXiv19,Scheibner_nHmech_PRL20}, electric circuits~\cite{Helbig_ExpSkin_19,Hofmann_ExpRecipSkin_19,Yoshida_MSkinPRR20}, and so on.

As well as the above dissipative systems, the non-Hermitian topology also provides a novel perspective of strongly correlated systems in equilibrium~\cite{VKozii_nH_arXiv17,Yoshida_EP_DMFT_PRB18}. 
In such a system, the single-particle spectrum is described by an effective non-Hermitian Hamiltonian which is composed of the non-interacting Hamiltonian and the self-energy~\cite{VKozii_nH_arXiv17,Yoshida_SPERs_PRB19,HShen2018quantum_osci,Zyuzin_nHEP_PRB18,Yoshida_EP_DMFT_PRB18,Papaji_nHEP_PRB19,Kimura_SPERs_PRB19,Matsushita_ER_PRB19,Michishita_EP_PRB20,Michishita_EP_PRL20,Rausch_EP1D_arXiv20,Matsushita_nHResp_arXiv20,Yoshida_nHReview_PTEP20}.
So far, it has been elucidated that exceptional points induce the bulk Fermi arc in the single-particle spectrum for equilibrium correlated systems.

In spite of the above progress of non-Hermitian band touching in the bulk, other topological properties inducing the skin effect have not been sufficiently explored in correlated systems.
While Okuma and Sato proposed that the topological properties of the skin effect are accessible by the pseudo-spectrum~\cite{Okuma_PS_aXiv20208}, an explicit numerical simulation of correlated systems is still missing.
In particular, it remains unclear in which correlated system the effective Hamiltonian possesses topological properties of the skin effect.

In this paper, by employing the real-space dynamical mean-field theory (R-DMFT), we analyze a two-dimensional correlated lattice model in equilibrium with special emphasis on the non-Hermitian topology of the skin effect.
Our R-DMFT simulation demonstrates that the non-trivial point-gap topology of the skin effect induces additional pseudo-eigenstates in contrast to the dependence of the density of states (DOS) on the boundary condition [for definition of the point-gap, see Eq.~(\ref{eq: def of point-gap})].
We also elucidate how the line-gap topology, another type of non-Hermitian topology, affects the pseudo-spectrum [for definition of the line-gap, see Eq.~(\ref{eq: def of line-gap})]. 
Our analysis clarifies that the damping of the quasi-particles has two effects; it induces the non-trivial point-gap topology and destroys the non-trivial line-gap topology. 
The above two effects result in distinct temperature dependences of the local pseudo-spectral weight depending on the momentum $k_x$; 
for $k_x=0$, the temperatures suppress the dependence of the local pseudo-spectral weight on the boundary conditions in contrast to the case for $k_x=\pi/2$.

The rest of this paper is organized as follows.
In Sec.~\ref{sec: model_rDMFT+IPT_PS}, we define the correlated model where the effective Hamiltonian shows the skin effect.
In this section, we also describes the framework of the R-DMFT and how to numerically extract the pseudo-spectrum from the R-DMFT data.
The R-DMFT results are presented in Sec.~\ref{sec: results all}.
In Appendices~\ref{sec: ps gen app}~and~\ref{sec: ps app}, we briefly review several properties of the pseudo-spectrum.

\section{Model and Method}
\label{sec: model_rDMFT+IPT_PS}

\subsection{Two-orbital Hubbard model}
\label{sec: model}

We analyze the following Hamiltonian
\begin{eqnarray}
\label{eq: model}
H &=& \sum_{\langle ij\rangle \alpha \beta \sigma}  h_{i\alpha j\beta} c^{\dagger}_{i\alpha\sigma} c_{j\beta\sigma} \nonumber \\
  && \quad\quad  + U\sum_{i} \left(n_{ib\uparrow}-\frac{1}{2}\right) \left(n_{ib\downarrow}-\frac{1}{2}\right),
\end{eqnarray}
where $c^{\dagger}_{j\beta\sigma}$ ($c_{j\beta\sigma}$) creates (annihilates) a fermion with a spin state $\sigma=\uparrow,\downarrow$ in orbital $\alpha=a,b$ at site $i$.
The matrix element $h_{i\alpha j\beta}\in \mathbb{C}$ is defined such that the Bloch Hamiltonian is written as
\begin{eqnarray}
\hat{h}(\bm{k}) &=& 
t(2\sin k_x -0.5\sin k_y) \hat{\tau}_2 +t(2\cos k_y) \hat{\tau}_3.
\end{eqnarray}
The second term of Eq.~(\ref{eq: model}) describes the repulsive interaction ($U\geq  0$); the number operator for fermions is defined as $n_{i\alpha\sigma}:=c^\dagger_{i\alpha\sigma} c _{i\alpha\sigma}$.

In this paper, we impose the periodic boundary condition along the $x$-direction (xPBC).
Along the $y$-direction, we impose either open or periodic boundary conditions. The former (latter) is denoted by yOBC (yPBC).
We note that $t$ is taken as the energy unit ($t=1$).

\subsection{R-DMFT}
\label{sec: rDMFT+IPT}
Here, we briefly describe the R-DMFT, an extended version of the dynamical mean-field theory~\cite{WMetznerPRL89_DMFT,MHartmannZP89_DMFT,AGeorgesRMP96_DMFT}, which allows us to treat systems with boundaries.
To be concrete, let us impose the xPBC and the yOBC on the model defined in Eq.~(\ref{eq: model}).
Here, $i_y$ ($i_y=0,1,...,L_y-1$) labels the sites along the $y$-direction.
In such a case, the system is inhomogeneous along the $y$-direction.

Within the R-DMFT framework, such inhomogeneity is encoded into the site-dependent self-energy $\Sigma_{i_y b \sigma}$ which is computed by mapping the lattice model to a set of effective impurity models. The impurity model specified by $i_y$ is written as
\begin{subequations}
\label{eq: eff imp}
\begin{eqnarray}
 \mathcal{Z}_{\mathrm{imp},i_y} &=& \int \!\! \mathcal{D}\bar{d}_{i_yb\sigma} \mathcal{D}d_{i_yb\sigma} e^{-S_{\mathrm{imp},i_y}},\\
 S_{\mathrm{imp},i_y} &=& -\int^\beta_0 \!\! d \tau d \tau'  \bar{d}_{i_yb\sigma}(\tau) \mathcal{G}^{-1}_{i_yb\sigma}(\tau-\tau')  d_{i_yb\sigma}(\tau') \nonumber \\
 &&\quad\quad \quad + \int^\beta_0 \!\! d \tau H_{\mathrm{imp},i_y}\delta(\tau), \\
 H_{\mathrm{imp},i_y} &=& U \left(n_{i_yb\uparrow}-\frac{1}{2}\right) \left(n_{i_yb\downarrow}-\frac{1}{2}\right).
\end{eqnarray}
\end{subequations}
Here, $\mathcal{Z}_{\mathrm{imp},i_y}$, $S_{\mathrm{imp},i_y}$, and $H_{\mathrm{imp},i_y}$ denote a partition function, an effective action, and a local Hamiltonian of each impurity problem, respectively.
The Grassmannian variables are denoted by $\bar{d}_{i_yb\sigma}$ and $d_{i_yb\sigma}$.

The Green's functions of the effective bath $\mathcal{G}_{i_yb\sigma}$ and the self-energy $\Sigma_{i_yb\sigma}$ are obtained by self-consistently solving the above impurity models~(\ref{eq: eff imp}) and the following equations:
\begin{subequations}
\begin{eqnarray}
  \mathcal{G}^{-1}_{i_yb\sigma} &=& G^{-1}_{i_yb\sigma} +\Sigma_{i_yb \sigma}, \\
  G_{i_yb\sigma} &=& \frac{1}{\sqrt{L_x}} \left[ \sum_{k_x} \left( (\omega+i\delta) \1 -\hat{h}(k_x) -\hat{\Sigma}_\sigma(\omega) \right) ^{-1} \right]_{i_yb i_yb}, \nonumber \\
\end{eqnarray}
with
\begin{eqnarray}
\label{eq: Sig_mat}
  \hat{\Sigma}_\sigma(\omega) &=& \mathrm{diag}\left( 0, \Sigma_{0b\sigma}(\omega+i\delta), 0, \Sigma_{1b\sigma}(\omega+i\delta), ...  \right). \nonumber \\
\end{eqnarray}
\end{subequations}
Here, $\hat{h}(k_x)$ is the Fourier transformed Hamiltonian along the $x$-direction for $U=0$, and $L_x$ denotes the number of sites along the $x$-direction.

In order to compute the self-energy of effective impurity models, we have employed an iterative perturbation theory~\cite{Georges_IPT_PRB92,Zhang_IPT_PRL93,Kajueter_modIPT_96} (IPT) based solver.

\subsection{Extracting pseudo-spectrum from R-DMFT data}
\label{sec: ps method}
The effective non-Hermitian Hamiltonian $\hat{H}_{\mathrm{eff}}(\omega,k_x):=\hat{h}(k_x)+\hat{\Sigma}(\omega)$ with the self-energy [see Eq.~(\ref{eq: Sig_mat})] may exhibit a non-Hermitian skin effect induced by the topological properties.

Reference~\onlinecite{Okuma_PS_aXiv20208} has proposed that the topological properties inducing the skin effect are accessible by the $\epsilon$-pseudo-spectrum ($\epsilon>0$) which is relevant to angle-resolved photoemission spectroscopy measurements.
In this section, we define the $\epsilon$-pseudo-spectrum and describe how to compute it (for more details, see Appendices~\ref{sec: ps gen app}~and~\ref{sec: ps app}).

The $\epsilon$-pseudo-spectrum $\sigma_\epsilon[\hat{H}_{\mathrm{eff}}(\omega,k_y)]$ with $\epsilon>0$ of the $N\times N$ matrix $\hat{H}_{\mathrm{eff}}(\omega,k_y)$ is defined as a set of pseudo-eigenvalues $\eta \in \mathbb{C}$ satisfying
\begin{eqnarray}
\label{eq: defs of ps_pseigval}
\twonorm{ [\eta \1- \hat{H}_{\mathrm{eff}}(\omega,k_y) ]\bm{v} }  &<& \epsilon,
\end{eqnarray}
with a pseudo-eigenvector $\bm{v}\in \mathbb{C}^{N}$ satisfying $\twonorm{\bm{v}}=1$.
The symbol $|\!| \bm{v} |\!|_2 $ denotes the norm of a vector $\bm{v}$; $|\!| \bm{v} |\!|_2:=[\sum_j v_jv^*_j]^{1/2}$ with $v^*_j$ being complex conjugation of the $j$-th component of the vector $\bm{v}$.

For computation of $\epsilon$-pseudo-spectrum, applying the singular value decomposition is useful; the $\epsilon$-pseudo-spectrum can be computed by searching pseudo-eigenvalues $\eta \in \mathbb{C}$ satisfying
\begin{eqnarray}
\label{eq: defs of ps svd}
s_{\mathrm{min}}[\eta \1- \hat{H}_{\mathrm{eff}}(\omega,k_x) ] &<&\epsilon,
\end{eqnarray}
where $s_{\mathrm{min}}(\hat{H}_{\mathrm{eff}})$ denotes the minimum singular value of a non-Hermitian matrix $\hat{H}_{\mathrm{eff}}$.
The pseudo-eigenvector is obtained from the unitary matrix of the singular value decomposition (see Appendix~\ref{sec: ps app}).

The above two definitions of the pseudo-spectrum are equivalent~\cite{Trefethen_psbook_2005}, which is shown in Appendix~\ref{sec: ps app}.

\section{
R-DMFT results
}
\label{sec: results all} 

In this section, after a brief discussion of the DOS, we see that $\hat{H}_{\mathrm{eff}}$ shows the skin effect because of the non-trivial point-gap topology [for definition of the point-gap, see Eq.~(\ref{eq: def of point-gap})]. 
Our R-DMFT simulation demonstrates that the above point-gap topology induces additional pseudo-eigenstates for the yOBC.
We also elucidate that additional pseudo-eigenstates also emerge due to the non-trivial line-gap topology [for definition of the line-gap, see Eq.~(\ref{eq: def of line-gap})].

The above two types of gap are defined as follows.
The point-gap of given energy $E_{\mathrm{ref}}\in \mathbb{C}$  opens when
\begin{eqnarray}
\label{eq: def of point-gap}
\varepsilon_n -E_{\mathrm{ref}} &\neq & 0,
\end{eqnarray}
holds for $n=0,1,...,\mathrm{dim}\hat{H}_{\mathrm{eff}}-1$. 
Here, $\varepsilon$'s are the eigenvalues of $\hat{H}_{\mathrm{eff}}(\omega=0,k_x)$.
The line-gap opens when 
\begin{eqnarray}
\label{eq: def of line-gap}
\mathrm{Re}\, \varepsilon_n &\neq& 0,
\end{eqnarray}
holds for $n=0,1,...,\mathrm{dim}\hat{H}_{\mathrm{eff}}-1$.

\subsection{
Density of states
}
\label{sec: free and LDOS} 
Let us start with the non-interacting case. As the Bloch Hamiltonian $\hat{h}(\bm{k})$ preserves the ``sublattice" symmetry~\cite{sublatsymm_ftnt} [$\hat{\tau}_1 \hat{h}(\bm{k}) \hat{\tau}_1 =-\hat{h}(\bm{k})$], the winding number may take a non-trivial value for a one-dimensional subsystem specified by $k_x$. 
Indeed, the winding number takes one for $-\pi \leq k_x \lesssim -2.88$, $-0.27 \lesssim k_x \lesssim 0.27$, and $-2.88 \lesssim k_x \leq  \pi$ otherwise it takes zero.
The non-trivial value of the winding number induces the zero energy states.

These zero energy states survive even in the presence of the Hubbard interaction $U$. 
The DOS $\rho(\omega,k_x)=\frac{1}{L_y}\sum_{i_y\alpha}[\hat{G}(\omega,k_x)]_{i_y\alpha i_y\alpha}$ is plotted in Figs.~\ref{fig: eigval-eigvec}(a)~and~\ref{fig: eigval-eigvec}(b).
For $k_x=0$ and the yOBC, 
the DOS shows a sharp peak at $\omega=0$ due to the edge states while such a peak cannot be observed for $k_x=\pi/2$.
As discussed in Sec.~\ref{sec: PS results}, the emergence of the sharp peak can be more clearly understood in terms of the non-Hermitian topology whose topological invariant is a non-Hermitian extension of the winding number [see Eq.~(\ref{eq: WL nH})].

\subsection{
Eigenvalues of the effective Hamiltonian
}
\label{sec: eigval_and_eigvec_Heff}

\begin{figure}[!t]
\begin{minipage}{1\hsize}
\begin{center}
\includegraphics[width=1\hsize,clip]{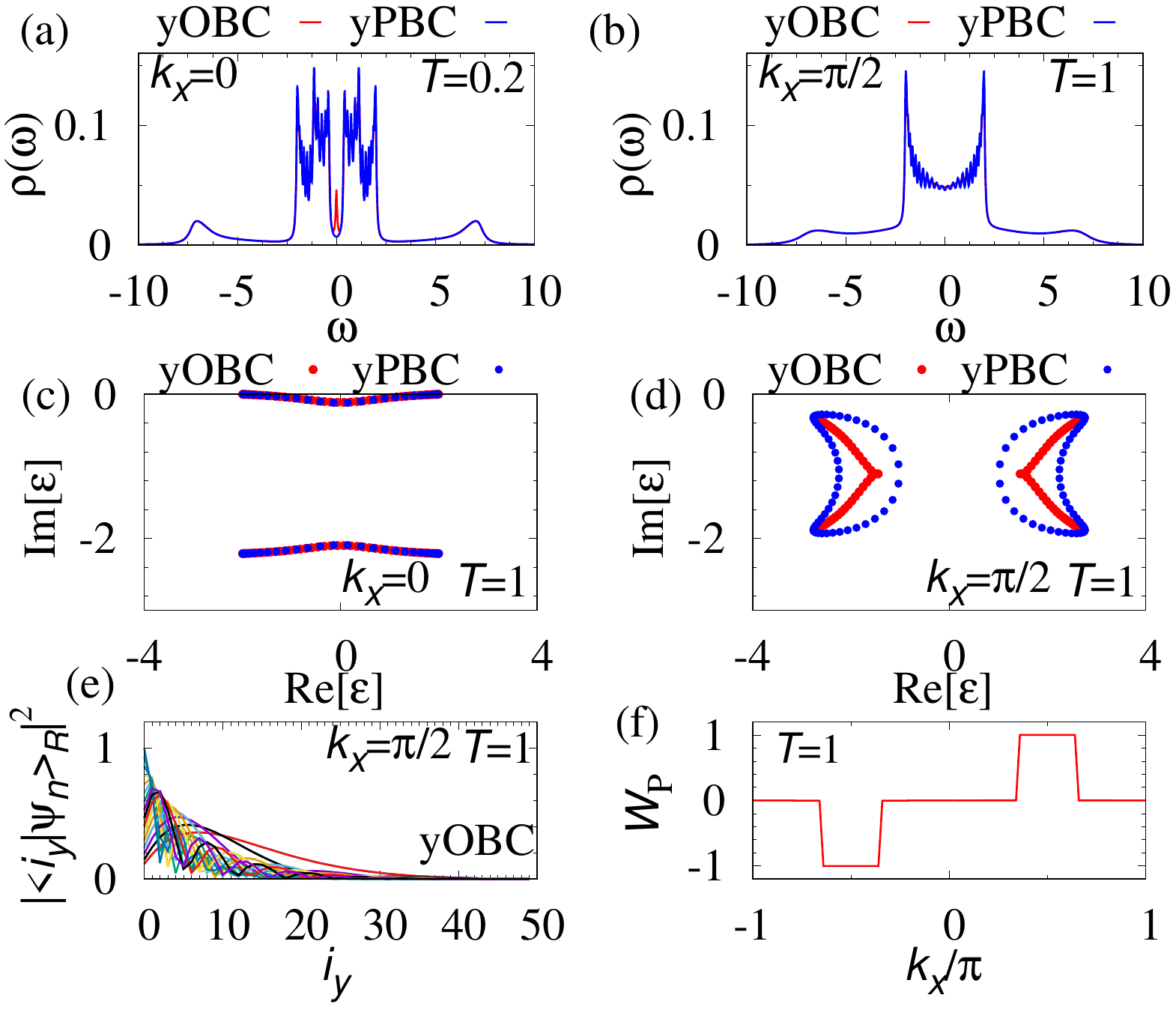}
\end{center}
\end{minipage}
\caption{
(a) [(b)]: Density of states $\rho(\omega,k_x)=\frac{1}{L_y}\sum_{i_y\alpha}[G(\omega,k_x)]_{i_y\alpha i_y\alpha}$ for $(k_x,T)=(0,0.2)$ [$(\pi/2,1)$].
(c) [(d)]: Spectrum of $\hat{H}_{\mathrm{eff}}(0,k_x)$ for $k_x=0$ [$\pi/2$]. The data represetned with red (blue) dots are obatined by imposing the yOBC (yPBC).
(e): Amplitude of right eigenvectors $|\langle i_y | \psi_n\rangle_R|$ for $k_x=\pi/2$. (f): Winding number $W_{\mathrm{P}}(k_x)$. 
Panels (c)-(f) are obtained for $T=1$.
}
\label{fig: eigval-eigvec}
\end{figure}

Now, we discuss whether the effective Hamiltonian $\hat{H}_{\mathrm{eff}}(\omega,k_x):=\hat{h}(k_x)+\hat{\Sigma}(\omega+i\delta)$ exhibits the skin effect, i.e., extreme sensitivity to the boundary conditions of the spectrum and eigenvectors.
As we see below, the effective Hamiltonian shows the skin effect for $k_x=\pi/2$.
Figures.~\ref{fig: eigval-eigvec}(c)~and~\ref{fig: eigval-eigvec}(d) show the eigenvalues of $\hat{H}_{\mathrm{eff}}(\omega=0,k_x)$ for $k_x=0$ and $k_x=\pi/2$, respectively.
These figures show that in contrast to the case for $k_x=0$, the spectrum for $k_x=\pi/2$ shows significant dependence of the boundary condition.
In addition, as shown in Fig.~\ref{fig: eigval-eigvec}(e), almost of all right eigenvectors $|\psi_{n}\rangle_R$ ($n=0,1,...,\mathrm{dim}\hat{H}_{\mathrm{eff}}-1$) are localized at $i_y=0$ for $k_x=\pi/2$ and the yOBC.

The above significant dependence of the spectrum is consistent with the non-trivial value of the winding number $W_{\mathrm{P}}(k_x,E_{\mathrm{ref}})$ of the point-gap;
\begin{eqnarray}
W_{\mathrm{P}}(k_x,E_{\mathrm{ref}}) &=& \int^\pi_{-\pi} \!\! \frac{dk_y}{2\pi i} \ \partial_{k_y} \log \mathrm{det}[\hat{H}_{\mathrm{eff}}(0,\bm{k})-E_{\mathrm{ref}}\1],\nonumber \\
\end{eqnarray}
where $E_{\mathrm{ref}}$ denotes the reference energy.
Figure~\ref{fig: eigval-eigvec}(f) indicates that the winding number takes one at $k_x=\pi/2$, where the skin effect is observed, while it takes zero at $k_x=0$.

The above results, (i.e., the significant dependence of the eigenvalues, the winding number $W_{\mathrm{P}}$, and the localization of eigenvectors), indicate that the effective Hamiltonian exhibits the skin effect for $(\omega,k_x)=(0,\pi/2)$.
The similar behaviors of the energy spectrum and the localization can be observed as long as the winding number takes a non-trivial value.

So far, we have seen that for $k_x=\pi/2$, the point-gap topology of the effective Hamiltonian $\hat{H}_{\mathrm{eff}}(\omega=0,k_x)$ becomes non-trivial and induces the skin effect [see Fig.~\ref{fig: eigval-eigvec}(c)-\ref{fig: eigval-eigvec}(f)], although 
the DOS does not show the corresponding extreme sensitivity to the boundary conditions [see Fig.~\ref{fig: eigval-eigvec}(b)].

\subsection{
Pseudo-spectrum
}
\label{sec: PS results}

\begin{figure*}[!t]
\begin{minipage}{1\hsize}
\begin{center}
\includegraphics[width=1\hsize,clip]{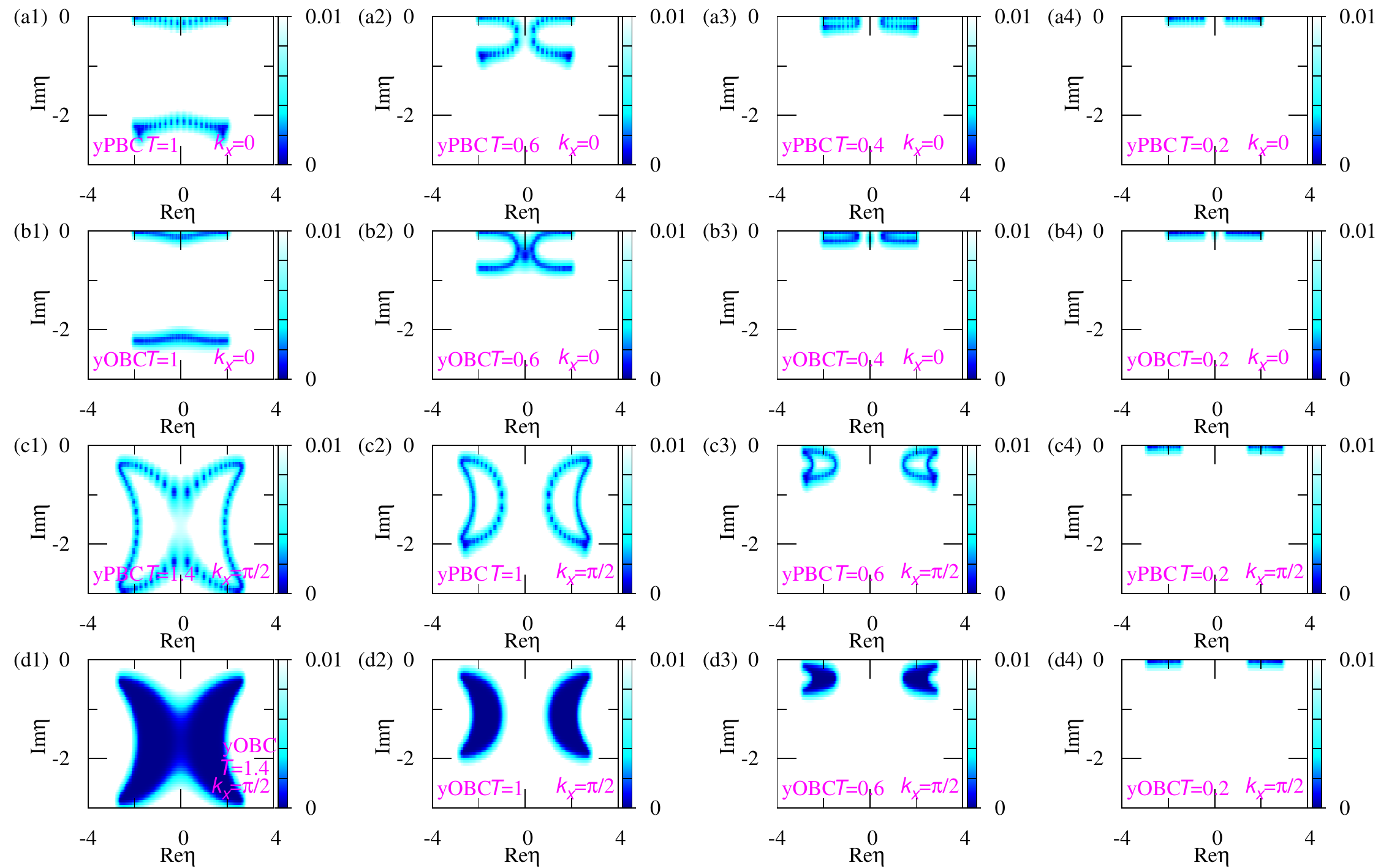}
\end{center}
\end{minipage}
\caption{
Pseudo-spectrum of the effective Hamiltonian $\hat{H}_{\mathrm{eff}}(\omega=0, k_x)$ for $L_y=50$ and $U=8$.
The color denotes the minimum singular value for a given $\eta$.
Panels (a1)-(a4) and (b1)-(b4) [(c1)-(c4) and (d1)-(d4)] are the data for $k_x=0$ ($k_x=\pi/2$).
Panels (a1)-(a4) and (c1)-(c4) [(b1)-(b4) and (d1)-(d4)] are obtained by imposing the yPBC (yOBC).
The data shown in panels (a1)-(a4) and (b1)-(b4) [(c1)-(c4) and (d1)-(d4)] are obtained for $T=1$, $0.6$, $0.4$, and $0.2$ [$T=1.4$, $1$, $0.6$, and $0.2$].
}
\label{fig: ps}
\end{figure*}

As proposed by Ref.~\onlinecite{Okuma_PS_aXiv20208}, the non-trivial value of the winding number $W_{\mathrm{P}}$, characterizing the skin effect, is reflected in the the pseudo-spectrum.
Figure~\ref{fig: ps} shows pseudo-spectrum of $\hat{H}_{\mathrm{eff}}(\omega=0,k_x)$.
Figures~\ref{fig: ps}(a1)-\ref{fig: ps}(a4)~and~\ref{fig: ps}(b1)-\ref{fig: ps}(b4) are the data for $k_x=0$ which we discuss the latter half of this section.

Figures~\ref{fig: ps}(c1)-\ref{fig: ps}(c4)~and~\ref{fig: ps}(d1)-\ref{fig: ps}(d4) indicate that the point-gap topology induces an additional structure of the pseudo-spectrum for the yOBC.
In order to see this, let us note that for $k_x=\pi/2$, the winding number $W_{\mathrm{P}}(\pi/2,E_{\mathrm{ref}})$ with $E_{\mathrm{ref}}=-2+0.5\times \Sigma_{\frac{L_y}{2}b\uparrow}(\omega=0)$ takes one for $0<T\lesssim0.09$. 
Figures~\ref{fig: ps}(c1)-\ref{fig: ps}(c4) show the pseudo-spectrum for $k_x=\pi/2$ and the yPBC.
These figures indicate that for the yPBC, the pseudo-spectrum shows a loop as the spectrum of $\hat{H}_{\mathrm{eff}}$ does.
For the yOBC, the pseudo-spectrum shows additional structures; besides the loop structure observed for the yPBC, the area enclosed by the loops also becomes pseudo-eigenvalues [see Fig.~\ref{fig: ps}(d1)-\ref{fig: ps}(d4)].
The above data indicate that the non-trivial topology of the point-gap induces the edge states of the pseudo-spectrum.

We recall that the non-trivial point-gap topology is induced by finite temperatures; finite temperature results in a finite lifetime of quasi-particles which is an origin of the loop structure of $\hat{H}_{\mathrm{eff}}(\omega=0,k_x)$ [see Fig.~\ref{fig: eigval-eigvec}(d)~and~\ref{fig: eigval-eigvec}(f)]. 

Besides the point-gap, one may find a line-gap, another type of gaps for non-Hermitian systems.
Our numerical data show that the non-trivial topology of the line-gap also affects pseudo-spectrum [see Figs.~\ref{fig: ps}(a1)-\ref{fig: ps}(a4)~and~\ref{fig: ps}(b1)-\ref{fig: ps}(b4)].

To see this, firstly, we introduce the winding number $W_\mathrm{L}(k_x)$ of the line-gap.
Because the non-Hermitian Hamiltonian preserves the ``sublattice" symmetry $\hat{\tau}_1 \hat{H}_{\mathrm{eff}}(\omega,\bm{k})\hat{\tau}_1 = -\hat{H}_{\mathrm{eff}}(\omega,\bm{k})$, the following winding number 
\begin{eqnarray}
\label{eq: WL nH}
W_\mathrm{L}(k_x) &=& \int^\pi_{-\pi} \!\! \frac{dk_y}{4\pi i} \ \mathrm{tr}[\hat{\tau}_1 \hat{H}^{-1}_{\mathrm{eff}}(0,\bm{k}) \partial_{k_y}   \hat{H}_{\mathrm{eff}}(0,\bm{k}) ],
\end{eqnarray}
is quantized as long as the line-gap opens.
For instance, $W_\mathrm{L}(0)$ takes one for $T\lesssim0.6$ and $U=8$, while $W_\mathrm{L}(\pi/2)$ takes zero for $T\lesssim 0.06$ and $U=8$.
The non-trivial line-gap topology for $k_x=0$ is reflected in the pseudo-spectrum.
Figures~\ref{fig: ps}(a1)-\ref{fig: ps}(a4) [\ref{fig: ps}(b1)-\ref{fig: ps}(b4)] show pseudo-spectrum for $k_x=0$ and the yPBC [yOBC].
In the low temperature region, $T\lesssim 0.6$, the pseudo-spectrum shows a peak on the imaginary axis $\mathrm{Re}\, \eta=0$ only for the yOBC [see Figs.~\ref{fig: ps}(b3)~and~\ref{fig: ps}(b4)], which corresponds to the non-trivial topology of $W_{\mathrm{L}}(0)$.
Increasing the temperature closes the line-gap, and the peak located on the imaginary axis disappears [see Figs.~\ref{fig: ps}(b1)~and~\ref{fig: ps}(b2)].
The above facts mean that additional pseudo-eigenenergy is induced by the non-trivial line-gap topology which are destroyed by increasing the temperature.

So far, we have seen that temperatures affect the two types of topology in the opposite way; increasing the temperature makes the point-gap topology non-trivial, in contrast, it destroys the non-trivial topology of the line-gap.
\begin{figure}[t]
\begin{minipage}{1\hsize}
\begin{center}
\includegraphics[width=1\hsize,clip]{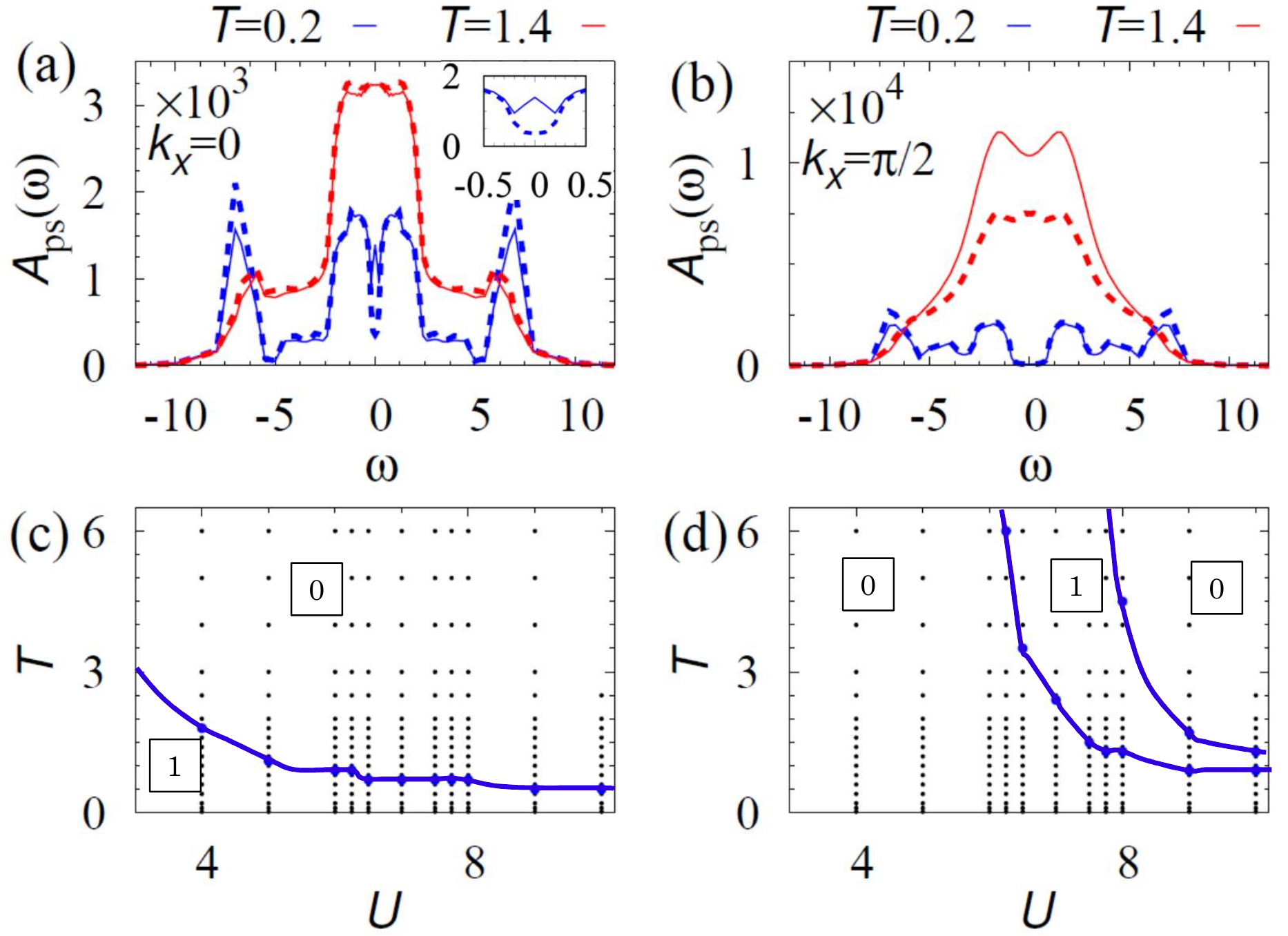}
\end{center}
\end{minipage}
\caption{
(a) and (b): The local pseudo-spectral weight for several cases of the boundary conditions and temperatures.
Panel (a) [(b)] shows the data for $k_x=0$ [$k_x=\pi/2$], respectively. These data are obtained for $U=8$.
(c) and (d): Phase diagrams of the interaction $U$ vs. the temperature $T$.
In panel (c) [(d)], numbers in squares indicate the values of $W_{\mathrm{L}}(k_x=0)$ [$W_{\mathrm{P}}(k_x,E_{\mathrm{ref}})$ with $k_x=\frac{\pi}{2}$ and $E_{\mathrm{ref}}=-2+0.5\times\Sigma_{\frac{L_y}{2}b\uparrow}(i\delta)$].
}
\label{fig: Aps_phase}
\end{figure}
This fact is also reflected in the following quantity
\begin{eqnarray}
A_{\mathrm{ps}}(\omega)&=& \sum_{\{ \eta \}} \frac{1}{\omega+i\delta - \eta },
\end{eqnarray}
where the summation is taken over all pseudo-eigenvalues. 
We call $A_{\mathrm{ps}}(\omega)$ the local pseudo-spectral weight because it is analogues to the local spectral weight~\cite{Schurcht_LSW_JoP2012,LSW_ftnt}.
In Figs.~\ref{fig: Aps_phase}(a)~and~\ref{fig: Aps_phase}(b), the local pseudo-spectral weight is plotted for several values of the temperature.
Figure~\ref{fig: Aps_phase}(a) indicates that only for low temperatures (e.g., $T=0.01$), $A_{\mathrm{ps}}$ at $k_x=0$ changes depending on the boundary conditions, which is no longer observed for $T=0.07$.
Figure~\ref{fig: Aps_phase}(b) indicates that for high temperatures (e.g., $T=0.07$) $A_{\mathrm{ps}}$ at $k_x=\pi/2$ changes depending on the boundary conditions.

The above temperature effects are summarized in Figs.~\ref{fig: Aps_phase}(c)~and~\ref{fig: Aps_phase}(d); the non-trivial topology of the line-gap (the point-gap) is destroyed by increasing (decreasing) the temperature.

\section{
Summary
}
\label{sec: summary}
By making use of the R-DMFT, we have analyzed the two-dimensional correlated system in equilibrium whose effective Hamiltonian exhibits the skin effect due to the non-trivial point-gap topology.
As well as the DOS, we have also computed the pseudo-spectrum which is recently proposed in Ref.~\onlinecite{Okuma_PS_aXiv20208}.

Our R-DMFT simulation has demonstrated that the above non-trivial point-gap topology induces additional pseudo-eigenstates for the yOBC in contrast to the dependence of the DOS on the boundary condition.
To our best knowledge, these are the first non-perturbative results elucidating the effects of the non-trivial topology of the skin effect on the spectral properties of correlated systems in equilibrium. 
We have further elucidated effects of the line-gap topology on the pseudo-spectrum. 
Our R-DMFT study has elucidated that in contrast to the point-gap topology, the damping of quasi-particles are harmful for the line-gap topology, which is reflected in temperature dependence of the local pseudo-spectral weight.

We consider that numerical simulations based on more accurate methods (e.g., methods based on the cluster DMFT~\cite{Kotliar_CDMFT_PRL01,Maier_CDMFT_RMP05} or a more accurate impurity solver such as numerical renormalization group~\cite{KWilsonRMP75_NRG,RPetersPRB06_NRG,RBullaRMP08_NRG} or continuous-time quantum Monte Carlo method~\cite{Werner_CTQMC_PRL06,Werner_CTQMC_PRB06,Haule_CTQMC_PRB07}) should yield the essentially same results because the above non-Hermitian topological properties are independent of quantitative details of the self-energy.

\section*{
Acknowledgements
}
This work is supported by JSPS Grant-in-Aid for Scientific Research on Innovative Areas ``Discrete Geometric Analysis for Materials Design": Grants No.~JP20H04627. 
This work is also supported by JSPS KAKENHI Grants No.~19K21032.
The author thanks the Supercomputer Center, the Institute for Solid State Physics, University of Tokyo for the use of the facilities.

\input{./ps_DMFT.bbl}

\appendix

\section{
Remarks of the pseudo-spectrum
}
\label{sec: ps gen app}

The following facts are discussed in Ref.~\onlinecite{Trefethen_psbook_2005}. However, we summarize them in order to make the paper self-contained.

\subsection{
Three definitions of the pseudo-spectrum
}
\label{sec: ps other defs app}
Consider an $N\times N$ matrix $\hat{A}\in \mathbb{C}^{N\times N}$. Then, there exist three definitions of the $\epsilon$-pseudo-spectra $\sigma_\epsilon(\hat{A})$ with $\epsilon>0$.
Before the details, we define a 2-norm of a matrix
\begin{eqnarray}
\label{eq: defs of twonorm of a matrix app}
\twonorm{\hat{A}}&=& 
\mathrm{max}_{\bm{v}} \frac{ \twonorm{ \hat{A} \bm{v} } }{ \twonorm{ \bm{v} } }.
\end{eqnarray}
We recall that $|\!| \bm{v} |\!|_2 $ denotes the norm of a vector $\bm{v}$; $|\!| \bm{v} |\!|_2:=[\sum_j v_jv^*_j]^{1/2}$ with $v^*_j$ being complex conjugation of the $j$-th component of the vector $\bm{v}$.

In the following, we see the three definitions of the $\epsilon$-pseudo-spectrum.

\textit{The first definition.--} The $\epsilon$-pseudo-spectrum is a set of pseudo-eigenvalues $\eta \in \mathbb{C}$ satisfying 
\begin{eqnarray}
\label{eq: defs of ps twonorm app}
{}|\!| [\eta \1- \hat{A} ]^{-1} |\!|_2  &>& \epsilon^{-1}.
\end{eqnarray}

\textit{The second definition.--} The $\epsilon$-pseudo-spectrum is the set of eigenvalues $\eta \in \mathbb{C}$ of the matrix $\hat{A}+ \hat{E}$ defined by a matrix $ \hat{E} \in \mathbb{C}^{N\times N}$ with $\twonorm{\hat{E}}<\epsilon$.
Namely, $\eta \in \mathbb{C}$ satisfies
\begin{eqnarray}
\label{eq: defs of ps eigval app}
(\hat{A}+ \hat{E})\bm{v} = \eta \bm{v},
\end{eqnarray}
with the eigenvector $\bm{v}$ ($\twonorm{\bm{v}}=1$).

\textit{The third definition.--} 
This equation is the same as the one defined in Eq.~(\ref{eq: defs of ps_pseigval}); the $\epsilon$-pseudo-spectrum is the set of $\eta \in \mathbb{C}$ satisfying
\begin{eqnarray}
\label{eq: defs of ps_pseigval app}
\twonorm{(\eta \1 -\hat{A})\bm{v}}<\epsilon,
\end{eqnarray}
for a normalized vector $\bm{v}$ ($\bm{v}\in\mathbb{C}^{N}$ and $\twonorm{\bm{v}}=1$).

In the next section, we prove the equivalence of these three definitions.

\subsection{
Equivalence of the three definitions
}
\label{sec: ps equiv other defs app}
Based on the argument provided in Ref.~\onlinecite{Trefethen_psbook_2005} (in particular, see page 16 of the textbook), we prove the equivalence of the above three definitions. 

In the following, we address the proof by the following three steps.

(i) Eq.~(\ref{eq: defs of ps eigval app}) $\Rightarrow $ Eq.~(\ref{eq: defs of ps_pseigval app})--.
Supposing that Eq.~(\ref{eq: defs of ps eigval app}) holds, we have Eq.~(\ref{eq: defs of ps_pseigval app}), which can be seen as follows.

Equation~(\ref{eq: defs of ps eigval app}) can be rewritten as
\begin{eqnarray}
(\eta\1 -\hat{A})\bm{v}&=& \hat{E}\bm{v}.
\end{eqnarray}
We note that the inequality $\twonorm{\hat{E}\bm{v}} < \epsilon $  holds because $\hat{E}$ satisfies $\twonorm{\hat{E}}<\epsilon$.
Thus, we obtain
\begin{eqnarray}
\twonorm{ (\eta\1-\hat{A})\bm{v} }&<& \epsilon,
\end{eqnarray}
with a vector $\bm{v}$ satisfying $\twonorm{\bm{v}}=1$.

(ii) Eq.~(\ref{eq: defs of ps_pseigval app}) $\Rightarrow $ Eq.~(\ref{eq: defs of ps twonorm app})--.
Supposing that Eq.~(\ref{eq: defs of ps_pseigval app}) holds, we have Eq.~(\ref{eq: defs of ps twonorm app}), which can be seen as follows.
Firstly, we note that Eq.~(\ref{eq: defs of ps_pseigval app}) indicates the relation
\begin{eqnarray}
\twonorm{(\eta \1 -\hat{A})\bm{v}} &=& s<\epsilon 
\end{eqnarray}
with $\bm{v}=1$ and an non-negative number $s\in \mathbb{R}$.
Thus, with a vector $\bm{u}\in \mathbb{C}^N$ satisfying $\twonorm{\bm{u}}=1$, we have 
\begin{eqnarray}
(\eta \1 -\hat{A})\bm{v} &=& s\bm{u},
\end{eqnarray}
which is further rewritten as
\begin{eqnarray}
(\eta \1-\hat{A})^{-1} \bm{u}&=& s^{-1}\bm{v}.
\end{eqnarray}
Recalling Eq.~(\ref{eq: defs of twonorm of a matrix app}), we have 
\begin{eqnarray}
\twonorm{(\eta\1-\hat{A})^{-1}}&\geq & \twonorm{(\eta\1-\hat{A})^{-1}\bm{u}}=s^{-1}>\epsilon^{-1}. \nonumber \\
\end{eqnarray}
Therefore, we obtain Eq.~(\ref{eq: defs of ps_pseigval app}) from Eq.~(\ref{eq: defs of ps twonorm app}).

(iii) Eq.~(\ref{eq: defs of ps twonorm app}) $\Rightarrow$ Eq.~(\ref{eq: defs of ps eigval app})--.
Supposing that Eq.~(\ref{eq: defs of ps twonorm app}) holds, we have Eq.~(\ref{eq: defs of ps eigval app}), which can be seen as follows.

Because Eq.~(\ref{eq: defs of twonorm of a matrix app}) holds, there exists $\bm{u}\in \mathbb{C}^N$ satisfying
\begin{eqnarray}
\twonorm{(\eta \1- \hat{A})}&=& \twonorm{(\eta \1- \hat{A})^{-1}\bm{u}} = s^{-1} >\epsilon^{-1},
\end{eqnarray}
which results in
\begin{eqnarray}
{} (\eta \1- \hat{A})^{-1}\bm{u} &=& s^{-1}\bm{v}.
\end{eqnarray}
This relation is further rewritten as
\begin{eqnarray}
{} (\eta \1- \hat{A})\bm{v} &=& s\bm{u}.
\end{eqnarray}
Because $s\bm{u}=\hat{E}\bm{v}$ ($\hat{E}:=s\bm{u}\bm{v}^\dagger$) holds, we have 
\begin{eqnarray}
{} (\eta \1- \hat{A})\bm{v} &=& \hat{E} \bm{v}.
\end{eqnarray}
Noting that $\twonorm{\hat{E}}=\twonorm{s\bm{u}\bm{v}^\dagger}=s<\epsilon$, we obtain Eq.~(\ref{eq: defs of ps eigval app}).

Putting the above arguments (i)-(iii) together, we end up with the equivalence of the three definitions.

\section{
Proof of Eq.~(\ref{eq: defs of ps svd})
}
\label{sec: ps app}
\subsection{Eq.~(\ref{eq: defs of ps_pseigval}) $\Leftrightarrow  $ Eq.~(\ref{eq: defs of ps svd}) }
Supposing that Eq.~(\ref{eq: defs of ps_pseigval}) holds, we have Eq.~(\ref{eq: defs of ps svd}), which can be seen as follows.

Firstly, we apply the singular value decomposition
\begin{eqnarray}
\eta \1-\hat{H}_{\mathrm{eff}} 
&=& 
\hat{U} \hat{\Lambda} \hat{V},
\end{eqnarray}
where $\hat{U}$ and $\hat{V}$ are unitary matrices and $\hat{\Lambda}$ is a diagonal matrix whose diagonal elements $\lambda_i \in \mathbb{R}$ ($i=1,2,...,N$) are non-negative.

Thus, the left hand side of Eq.~(\ref{eq: defs of ps_pseigval}) is rewritten as
\begin{eqnarray}
\label{eq: |eta -H| svd}
\twonorm{ (\eta \1-\hat{H}_{\mathrm{eff}})\bm{v} } &=& (\bm{v}^\dagger  \hat{V}^\dagger \hat{\Lambda} \hat{U}^\dagger \hat{U} \hat{\Lambda} \hat{V}  \bm{v} )^{1/2} \nonumber \\
&=& (\bm{v}^\dagger  \hat{V}^\dagger \hat{\Lambda}^2 \hat{V}  \bm{v} )^{1/2} \nonumber \\
&=& (\sum_{j}  |v'_j|^2  \lambda^2_j )^{1/2},
\end{eqnarray}
where $v'_j$ is defined as $v'_j:=\sum_i V_{ji}v_i$.

The above facts indicate that $\eta$ is an $\epsilon$-pseudo-eigenvalue ($\epsilon>0$) when
\begin{eqnarray}
s_{\mathrm{min}}(\eta \1-\hat{H}_{\mathrm{eff}}) &<& \epsilon,
\end{eqnarray}
is satisfied. 
Here, $s_{\mathrm{min}}(\eta \1-\hat{H}_{\mathrm{eff}})$ denotes the minimum of the singular values $\lambda$'s. 

Therefore, we obtain Eq.~(\ref{eq: defs of ps svd}).
Equation~(\ref{eq: |eta -H| svd}) also indicates that the pseudo-eigenvector $\bm{v}$ is given by $v_j=(V^\dagger)_{j1}$.

\subsection{
Another proof
}

Equation~(\ref{eq: defs of ps svd}) can also be proven from Eq.~(\ref{eq: defs of twonorm of a matrix app}).
This can be seen by noticing the following fact:
for an arbitrary matrix $\hat{B}\in \mathbb{C}^{N\times N}$,
\begin{eqnarray}
\label{eq: |B|_2 and svd B app}
\twonorm{\hat{B}}&=& s_{\mathrm{max}}(\hat{B}),
\end{eqnarray}
where $s_{\mathrm{max}}(\hat{B})$ is the maximum singular value of an $N\times N$ matrix $\hat{B}\in \mathbb{C}^{N \times N}$.

With Eq.~(\ref{eq: |B|_2 and svd B app}), we obtain
\begin{eqnarray}
\twonorm{ [\eta \1-\hat{H}_{\mathrm{eff}}]^{-1} } &=&
s_{\mathrm{max}}([\eta \1-\hat{H}_{\mathrm{eff}}]^{-1}) \nonumber \\
&=&
[s_{\mathrm{min}}(\eta \1-\hat{H}_{\mathrm{eff}})]^{-1}.
\end{eqnarray}
Combining the above equation and Eq.~(\ref{eq: defs of ps twonorm app}) results in Eq.~(\ref{eq: defs of ps svd}).

In the following, we proof Eq.~(\ref{eq: |B|_2 and svd B app}).
Firstly, we note that with the singular value decomposition ($\hat{B}:=\hat{U}_B \hat{\Lambda}_B \hat{V}_B$), we have
\begin{eqnarray}
\label{eq: |Bv|_2 svd app}
\twonorm{\hat{B}\bm{v}}&=& \sqrt{ \bm{v}^\dagger \hat{B}^\dagger \hat{B}\bm{v} } \nonumber \\
                 &=& \sqrt{ \bm{v}^\dagger \hat{V}^\dagger_B \hat{\Lambda}_B  \hat{U}^\dagger_B \hat{U}_B \hat{\Lambda}_B \hat{V}_B \bm{v} } \nonumber \\
                 &=& \sqrt{ \bm{v}^\dagger \hat{V}^\dagger_B \hat{\Lambda}^2_B \hat{V}_B \bm{v} }.
\end{eqnarray}
Here $\hat{U}_B$ and $\hat{V}_B$ are unitary matrices, and $\hat{\Lambda}_B$ is a diagonal matrix whose diagonal elements $\lambda_{Bi}$ ($i=1,2,...,N$) are singular values of $\hat{B}$. 
The vector $\bm{v}$ is an arbitrary vector $\bm{v}\in \mathbb{C}^N$.

By making use of Eq.~(\ref{eq: |Bv|_2 svd app}), we obtain Eq.~(\ref{eq: |B|_2 and svd B app})
\begin{eqnarray}
\twonorm{\hat{B}} &=& \mathrm{max}_{\bm{n}} \twonorm{\hat{B}\bm{n}}\nonumber  \\
            &=& \mathrm{max}_{\bm{n}} \sqrt{ \bm{n}^\dagger \hat{V}^\dagger_B \hat{\Lambda}^2_B \hat{V}_B \bm{n} } \nonumber \\
            &=& \mathrm{max}_{\bm{n'}} \sqrt{ \bm{n}'^\dagger \hat{\Lambda}^2_B \bm{n}' } \nonumber \\
            &=& \mathrm{max}_{\bm{n'}} (\sum_j |n'|^2_{Bj}\lambda^2_{Bj})^{1/2} \nonumber \\
            &=& s_{\mathrm{max}}(\hat{B}),
\end{eqnarray}
where we have defined a vector $\bm{n}':=\hat{V}\bm{n}$ satisfying $\twonorm{\bm{n}'}=1$.
Thus, we obtain Eq.~(\ref{eq: |B|_2 and svd B app}).

As discussed above, we can also obtain Eq.~(\ref{eq: defs of ps svd}) starting from Eq.~(\ref{eq: defs of twonorm of a matrix app}).

\end{document}

%% file: ps_DMFT.bbl
%

%% file: ps_DMFT_1109_5_submit.bbl
\begin{thebibliography}{94}%
\makeatletter
\providecommand \@ifxundefined [1]{%
 \@ifx{#1\undefined}
}%
\providecommand \@ifnum [1]{%
 \ifnum #1\expandafter \@firstoftwo
 \else \expandafter \@secondoftwo
 \fi
}%
\providecommand \@ifx [1]{%
 \ifx #1\expandafter \@firstoftwo
 \else \expandafter \@secondoftwo
 \fi
}%
\providecommand \natexlab [1]{#1}%
\providecommand \enquote  [1]{``#1''}%
\providecommand \bibnamefont  [1]{#1}%
\providecommand \bibfnamefont [1]{#1}%
\providecommand \citenamefont [1]{#1}%
\providecommand \href@noop [0]{\@secondoftwo}%
\providecommand \href [0]{\begingroup \@sanitize@url \@href}%
\providecommand \@href[1]{\@@startlink{#1}\@@href}%
\providecommand \@@href[1]{\endgroup#1\@@endlink}%
\providecommand \@sanitize@url [0]{\catcode `\\12\catcode `\$12\catcode
  `\&12\catcode `\#12\catcode `\^12\catcode `\_12\catcode `\%12\relax}%
\providecommand \@@startlink[1]{}%
\providecommand \@@endlink[0]{}%
\providecommand \url  [0]{\begingroup\@sanitize@url \@url }%
\providecommand \@url [1]{\endgroup\@href {#1}{\urlprefix }}%
\providecommand \urlprefix  [0]{URL }%
\providecommand \Eprint [0]{\href }%
\providecommand \doibase [0]{http://dx.doi.org/}%
\providecommand \selectlanguage [0]{\@gobble}%
\providecommand \bibinfo  [0]{\@secondoftwo}%
\providecommand \bibfield  [0]{\@secondoftwo}%
\providecommand \translation [1]{[#1]}%
\providecommand \BibitemOpen [0]{}%
\providecommand \bibitemStop [0]{}%
\providecommand \bibitemNoStop [0]{.\EOS\space}%
\providecommand \EOS [0]{\spacefactor3000\relax}%
\providecommand \BibitemShut  [1]{\csname bibitem#1\endcsname}%
\let\auto@bib@innerbib\@empty
\bibitem [{\citenamefont {Kane}\ and\ \citenamefont
  {Mele}(2005{\natexlab{a}})}]{Kane_Z2TI_PRL05_1}%
  \BibitemOpen
  \bibfield  {author} {\bibinfo {author} {\bibfnamefont {C.~L.}\ \bibnamefont
  {Kane}}\ and\ \bibinfo {author} {\bibfnamefont {E.~J.}\ \bibnamefont
  {Mele}},\ }\href {\doibase 10.1103/PhysRevLett.95.146802} {\bibfield
  {journal} {\bibinfo  {journal} {Phys. Rev. Lett.}\ }\textbf {\bibinfo
  {volume} {95}},\ \bibinfo {pages} {146802} (\bibinfo {year}
  {2005}{\natexlab{a}})}\BibitemShut {NoStop}%
\bibitem [{\citenamefont {Kane}\ and\ \citenamefont
  {Mele}(2005{\natexlab{b}})}]{Kane_Z2TI_PRL05_2}%
  \BibitemOpen
  \bibfield  {author} {\bibinfo {author} {\bibfnamefont {C.~L.}\ \bibnamefont
  {Kane}}\ and\ \bibinfo {author} {\bibfnamefont {E.~J.}\ \bibnamefont
  {Mele}},\ }\href {\doibase 10.1103/PhysRevLett.95.226801} {\bibfield
  {journal} {\bibinfo  {journal} {Phys. Rev. Lett.}\ }\textbf {\bibinfo
  {volume} {95}},\ \bibinfo {pages} {226801} (\bibinfo {year}
  {2005}{\natexlab{b}})}\BibitemShut {NoStop}%
\bibitem [{\citenamefont {Bernevig}\ \emph {et~al.}(2006)\citenamefont
  {Bernevig}, \citenamefont {Hughes},\ and\ \citenamefont
  {Zhang}}]{HgTe_Bernevig06}%
  \BibitemOpen
  \bibfield  {author} {\bibinfo {author} {\bibfnamefont {B.~A.}\ \bibnamefont
  {Bernevig}}, \bibinfo {author} {\bibfnamefont {T.~L.}\ \bibnamefont
  {Hughes}}, \ and\ \bibinfo {author} {\bibfnamefont {S.-C.}\ \bibnamefont
  {Zhang}},\ }\href {\doibase 10.1126/science.1133734} {\bibfield  {journal}
  {\bibinfo  {journal} {Science}\ }\textbf {\bibinfo {volume} {314}},\ \bibinfo
  {pages} {1757} (\bibinfo {year} {2006})}\BibitemShut {NoStop}%
\bibitem [{\citenamefont {K\"onig}\ \emph {et~al.}(2007)\citenamefont
  {K\"onig}, \citenamefont {Wiedmann}, \citenamefont {Br\"une}, \citenamefont
  {Roth}, \citenamefont {Buhmann}, \citenamefont {Molenkamp}, \citenamefont
  {Qi},\ and\ \citenamefont {Zhang}}]{Konig_QSHE2007}%
  \BibitemOpen
  \bibfield  {author} {\bibinfo {author} {\bibfnamefont {M.}~\bibnamefont
  {K\"onig}}, \bibinfo {author} {\bibfnamefont {S.}~\bibnamefont {Wiedmann}},
  \bibinfo {author} {\bibfnamefont {C.}~\bibnamefont {Br\"une}}, \bibinfo
  {author} {\bibfnamefont {A.}~\bibnamefont {Roth}}, \bibinfo {author}
  {\bibfnamefont {H.}~\bibnamefont {Buhmann}}, \bibinfo {author} {\bibfnamefont
  {L.~W.}\ \bibnamefont {Molenkamp}}, \bibinfo {author} {\bibfnamefont {X.-L.}\
  \bibnamefont {Qi}}, \ and\ \bibinfo {author} {\bibfnamefont {S.-C.}\
  \bibnamefont {Zhang}},\ }\href {\doibase 10.1126/science.1148047} {\bibfield
  {journal} {\bibinfo  {journal} {Science}\ }\textbf {\bibinfo {volume}
  {318}},\ \bibinfo {pages} {766} (\bibinfo {year} {2007})}\BibitemShut
  {NoStop}%
\bibitem [{\citenamefont {Qi}\ \emph {et~al.}(2008)\citenamefont {Qi},
  \citenamefont {Hughes},\ and\ \citenamefont {Zhang}}]{Qi_TQFTofTI_PRB08}%
  \BibitemOpen
  \bibfield  {author} {\bibinfo {author} {\bibfnamefont {X.-L.}\ \bibnamefont
  {Qi}}, \bibinfo {author} {\bibfnamefont {T.~L.}\ \bibnamefont {Hughes}}, \
  and\ \bibinfo {author} {\bibfnamefont {S.-C.}\ \bibnamefont {Zhang}},\ }\href
  {\doibase 10.1103/PhysRevB.78.195424} {\bibfield  {journal} {\bibinfo
  {journal} {Phys. Rev. B}\ }\textbf {\bibinfo {volume} {78}},\ \bibinfo
  {pages} {195424} (\bibinfo {year} {2008})}\BibitemShut {NoStop}%
\bibitem [{\citenamefont {Hasan}\ and\ \citenamefont
  {Kane}(2010)}]{TI_review_Hasan10}%
  \BibitemOpen
  \bibfield  {author} {\bibinfo {author} {\bibfnamefont {M.~Z.}\ \bibnamefont
  {Hasan}}\ and\ \bibinfo {author} {\bibfnamefont {C.~L.}\ \bibnamefont
  {Kane}},\ }\href {\doibase 10.1103/RevModPhys.82.3045} {\bibfield  {journal}
  {\bibinfo  {journal} {Rev. Mod. Phys.}\ }\textbf {\bibinfo {volume} {82}},\
  \bibinfo {pages} {3045} (\bibinfo {year} {2010})}\BibitemShut {NoStop}%
\bibitem [{\citenamefont {Qi}\ and\ \citenamefont
  {Zhang}(2011)}]{TI_review_Qi10}%
  \BibitemOpen
  \bibfield  {author} {\bibinfo {author} {\bibfnamefont {X.-L.}\ \bibnamefont
  {Qi}}\ and\ \bibinfo {author} {\bibfnamefont {S.-C.}\ \bibnamefont {Zhang}},\
  }\href {\doibase 10.1103/RevModPhys.83.1057} {\bibfield  {journal} {\bibinfo
  {journal} {Rev. Mod. Phys.}\ }\textbf {\bibinfo {volume} {83}},\ \bibinfo
  {pages} {1057} (\bibinfo {year} {2011})}\BibitemShut {NoStop}%
\bibitem [{\citenamefont {Hatano}\ and\ \citenamefont
  {Nelson}(1996)}]{Hatano_PRL96}%
  \BibitemOpen
  \bibfield  {author} {\bibinfo {author} {\bibfnamefont {N.}~\bibnamefont
  {Hatano}}\ and\ \bibinfo {author} {\bibfnamefont {D.~R.}\ \bibnamefont
  {Nelson}},\ }\href {\doibase 10.1103/PhysRevLett.77.570} {\bibfield
  {journal} {\bibinfo  {journal} {Phys. Rev. Lett.}\ }\textbf {\bibinfo
  {volume} {77}},\ \bibinfo {pages} {570} (\bibinfo {year} {1996})}\BibitemShut
  {NoStop}%
\bibitem [{\citenamefont {Hu}\ and\ \citenamefont
  {Hughes}(2011)}]{Hu_nH_PRB11}%
  \BibitemOpen
  \bibfield  {author} {\bibinfo {author} {\bibfnamefont {Y.~C.}\ \bibnamefont
  {Hu}}\ and\ \bibinfo {author} {\bibfnamefont {T.~L.}\ \bibnamefont
  {Hughes}},\ }\href {\doibase 10.1103/PhysRevB.84.153101} {\bibfield
  {journal} {\bibinfo  {journal} {Phys. Rev. B}\ }\textbf {\bibinfo {volume}
  {84}},\ \bibinfo {pages} {153101} (\bibinfo {year} {2011})}\BibitemShut
  {NoStop}%
\bibitem [{\citenamefont {Esaki}\ \emph {et~al.}(2011)\citenamefont {Esaki},
  \citenamefont {Sato}, \citenamefont {Hasebe},\ and\ \citenamefont
  {Kohmoto}}]{Esaki_nH_PRB11}%
  \BibitemOpen
  \bibfield  {author} {\bibinfo {author} {\bibfnamefont {K.}~\bibnamefont
  {Esaki}}, \bibinfo {author} {\bibfnamefont {M.}~\bibnamefont {Sato}},
  \bibinfo {author} {\bibfnamefont {K.}~\bibnamefont {Hasebe}}, \ and\ \bibinfo
  {author} {\bibfnamefont {M.}~\bibnamefont {Kohmoto}},\ }\href {\doibase
  10.1103/PhysRevB.84.205128} {\bibfield  {journal} {\bibinfo  {journal} {Phys.
  Rev. B}\ }\textbf {\bibinfo {volume} {84}},\ \bibinfo {pages} {205128}
  (\bibinfo {year} {2011})}\BibitemShut {NoStop}%
\bibitem [{\citenamefont {Lee}(2016)}]{TELeePRL16_Half_quantized}%
  \BibitemOpen
  \bibfield  {author} {\bibinfo {author} {\bibfnamefont {T.~E.}\ \bibnamefont
  {Lee}},\ }\href {\doibase 10.1103/PhysRevLett.116.133903} {\bibfield
  {journal} {\bibinfo  {journal} {Phys. Rev. Lett.}\ }\textbf {\bibinfo
  {volume} {116}},\ \bibinfo {pages} {133903} (\bibinfo {year}
  {2016})}\BibitemShut {NoStop}%
\bibitem [{\citenamefont {Kunst}\ \emph {et~al.}(2018)\citenamefont {Kunst},
  \citenamefont {Edvardsson}, \citenamefont {Budich},\ and\ \citenamefont
  {Bergholtz}}]{KFlore_nHSkin_PRL18}%
  \BibitemOpen
  \bibfield  {author} {\bibinfo {author} {\bibfnamefont {F.~K.}\ \bibnamefont
  {Kunst}}, \bibinfo {author} {\bibfnamefont {E.}~\bibnamefont {Edvardsson}},
  \bibinfo {author} {\bibfnamefont {J.~C.}\ \bibnamefont {Budich}}, \ and\
  \bibinfo {author} {\bibfnamefont {E.~J.}\ \bibnamefont {Bergholtz}},\ }\href
  {\doibase 10.1103/PhysRevLett.121.026808} {\bibfield  {journal} {\bibinfo
  {journal} {Phys. Rev. Lett.}\ }\textbf {\bibinfo {volume} {121}},\ \bibinfo
  {pages} {026808} (\bibinfo {year} {2018})}\BibitemShut {NoStop}%
\bibitem [{\citenamefont {Edvardsson}\ \emph {et~al.}(2019)\citenamefont
  {Edvardsson}, \citenamefont {Kunst},\ and\ \citenamefont
  {Bergholtz}}]{EElizabet_PRBnHSkinHOTI_PRB19}%
  \BibitemOpen
  \bibfield  {author} {\bibinfo {author} {\bibfnamefont {E.}~\bibnamefont
  {Edvardsson}}, \bibinfo {author} {\bibfnamefont {F.~K.}\ \bibnamefont
  {Kunst}}, \ and\ \bibinfo {author} {\bibfnamefont {E.~J.}\ \bibnamefont
  {Bergholtz}},\ }\href {\doibase 10.1103/PhysRevB.99.081302} {\bibfield
  {journal} {\bibinfo  {journal} {Phys. Rev. B}\ }\textbf {\bibinfo {volume}
  {99}},\ \bibinfo {pages} {081302} (\bibinfo {year} {2019})}\BibitemShut
  {NoStop}%
\bibitem [{\citenamefont {Gong}\ \emph {et~al.}(2018)\citenamefont {Gong},
  \citenamefont {Ashida}, \citenamefont {Kawabata}, \citenamefont {Takasan},
  \citenamefont {Higashikawa},\ and\ \citenamefont {Ueda}}]{Gong_class_PRX18}%
  \BibitemOpen
  \bibfield  {author} {\bibinfo {author} {\bibfnamefont {Z.}~\bibnamefont
  {Gong}}, \bibinfo {author} {\bibfnamefont {Y.}~\bibnamefont {Ashida}},
  \bibinfo {author} {\bibfnamefont {K.}~\bibnamefont {Kawabata}}, \bibinfo
  {author} {\bibfnamefont {K.}~\bibnamefont {Takasan}}, \bibinfo {author}
  {\bibfnamefont {S.}~\bibnamefont {Higashikawa}}, \ and\ \bibinfo {author}
  {\bibfnamefont {M.}~\bibnamefont {Ueda}},\ }\href {\doibase
  10.1103/PhysRevX.8.031079} {\bibfield  {journal} {\bibinfo  {journal} {Phys.
  Rev. X}\ }\textbf {\bibinfo {volume} {8}},\ \bibinfo {pages} {031079}
  (\bibinfo {year} {2018})}\BibitemShut {NoStop}%
\bibitem [{\citenamefont {Kawabata}\ \emph
  {et~al.}(2019{\natexlab{a}})\citenamefont {Kawabata}, \citenamefont
  {Shiozaki}, \citenamefont {Ueda},\ and\ \citenamefont
  {Sato}}]{Kawabata_gapped_PRX19}%
  \BibitemOpen
  \bibfield  {author} {\bibinfo {author} {\bibfnamefont {K.}~\bibnamefont
  {Kawabata}}, \bibinfo {author} {\bibfnamefont {K.}~\bibnamefont {Shiozaki}},
  \bibinfo {author} {\bibfnamefont {M.}~\bibnamefont {Ueda}}, \ and\ \bibinfo
  {author} {\bibfnamefont {M.}~\bibnamefont {Sato}},\ }\href {\doibase
  10.1103/PhysRevX.9.041015} {\bibfield  {journal} {\bibinfo  {journal} {Phys.
  Rev. X}\ }\textbf {\bibinfo {volume} {9}},\ \bibinfo {pages} {041015}
  (\bibinfo {year} {2019}{\natexlab{a}})}\BibitemShut {NoStop}%
\bibitem [{\citenamefont {Zhou}\ and\ \citenamefont
  {Lee}(2019)}]{Zhou_gapped_class_PRB19}%
  \BibitemOpen
  \bibfield  {author} {\bibinfo {author} {\bibfnamefont {H.}~\bibnamefont
  {Zhou}}\ and\ \bibinfo {author} {\bibfnamefont {J.~Y.}\ \bibnamefont {Lee}},\
  }\href {\doibase 10.1103/PhysRevB.99.235112} {\bibfield  {journal} {\bibinfo
  {journal} {Phys. Rev. B}\ }\textbf {\bibinfo {volume} {99}},\ \bibinfo
  {pages} {235112} (\bibinfo {year} {2019})}\BibitemShut {NoStop}%
\bibitem [{\citenamefont {Yokomizo}\ and\ \citenamefont
  {Murakami}(2019)}]{Yokomizo_BBC_PRL19}%
  \BibitemOpen
  \bibfield  {author} {\bibinfo {author} {\bibfnamefont {K.}~\bibnamefont
  {Yokomizo}}\ and\ \bibinfo {author} {\bibfnamefont {S.}~\bibnamefont
  {Murakami}},\ }\href {\doibase 10.1103/PhysRevLett.123.066404} {\bibfield
  {journal} {\bibinfo  {journal} {Phys. Rev. Lett.}\ }\textbf {\bibinfo
  {volume} {123}},\ \bibinfo {pages} {066404} (\bibinfo {year}
  {2019})}\BibitemShut {NoStop}%
\bibitem [{\citenamefont {Okuma}\ and\ \citenamefont
  {Sato}(2019)}]{Okuma_BECpg_PRL19}%
  \BibitemOpen
  \bibfield  {author} {\bibinfo {author} {\bibfnamefont {N.}~\bibnamefont
  {Okuma}}\ and\ \bibinfo {author} {\bibfnamefont {M.}~\bibnamefont {Sato}},\
  }\href {\doibase 10.1103/PhysRevLett.123.097701} {\bibfield  {journal}
  {\bibinfo  {journal} {Phys. Rev. Lett.}\ }\textbf {\bibinfo {volume} {123}},\
  \bibinfo {pages} {097701} (\bibinfo {year} {2019})}\BibitemShut {NoStop}%
\bibitem [{\citenamefont {Yokomizo}\ and\ \citenamefont
  {Murakami}(2020)}]{Yokomizo_NBlochBBCEP_arXiv20}%
  \BibitemOpen
  \bibfield  {author} {\bibinfo {author} {\bibfnamefont {K.}~\bibnamefont
  {Yokomizo}}\ and\ \bibinfo {author} {\bibfnamefont {S.}~\bibnamefont
  {Murakami}},\ }\href@noop {} {\bibfield  {journal} {\bibinfo  {journal}
  {arXiv preprint arXiv:2001.07348}\ } (\bibinfo {year} {2020})}\BibitemShut
  {NoStop}%
\bibitem [{\citenamefont {Bergholtz}\ \emph {et~al.}(2019)\citenamefont
  {Bergholtz}, \citenamefont {Budich},\ and\ \citenamefont
  {Kunst}}]{Bergholtz_Review19}%
  \BibitemOpen
  \bibfield  {author} {\bibinfo {author} {\bibfnamefont {E.~J.}\ \bibnamefont
  {Bergholtz}}, \bibinfo {author} {\bibfnamefont {J.~C.}\ \bibnamefont
  {Budich}}, \ and\ \bibinfo {author} {\bibfnamefont {F.~K.}\ \bibnamefont
  {Kunst}},\ }\href@noop {} {\bibfield  {journal} {\bibinfo  {journal} {arXiv
  preprint arXiv:1912.10048}\ } (\bibinfo {year} {2019})}\BibitemShut {NoStop}%
\bibitem [{\citenamefont {Yoshida}\ \emph
  {et~al.}(2020{\natexlab{a}})\citenamefont {Yoshida}, \citenamefont {Peters},
  \citenamefont {Kawakami},\ and\ \citenamefont
  {Hatsugai}}]{Yoshida_nHReview_PTEP20}%
  \BibitemOpen
  \bibfield  {author} {\bibinfo {author} {\bibfnamefont {T.}~\bibnamefont
  {Yoshida}}, \bibinfo {author} {\bibfnamefont {R.}~\bibnamefont {Peters}},
  \bibinfo {author} {\bibfnamefont {N.}~\bibnamefont {Kawakami}}, \ and\
  \bibinfo {author} {\bibfnamefont {Y.}~\bibnamefont {Hatsugai}},\ }\href
  {\doibase 10.1093/ptep/ptaa059} {\bibfield  {journal} {\bibinfo  {journal}
  {Progress of Theoretical and Experimental Physics}\ } (\bibinfo {year}
  {2020}{\natexlab{a}}),\ 10.1093/ptep/ptaa059},\ \bibinfo {note} {ptaa059},\
  \Eprint
  {http://arxiv.org/abs/https://academic.oup.com/ptep/advance-article-pdf/doi/10.1093/ptep/ptaa059/33529446/ptaa059.pdf}
  {https://academic.oup.com/ptep/advance-article-pdf/doi/10.1093/ptep/ptaa059/33529446/ptaa059.pdf}
  \BibitemShut {NoStop}%
\bibitem [{\citenamefont {Ashida}\ \emph {et~al.}(2020)\citenamefont {Ashida},
  \citenamefont {Gong},\ and\ \citenamefont {Ueda}}]{Ashida_nHReview_arXiv20}%
  \BibitemOpen
  \bibfield  {author} {\bibinfo {author} {\bibfnamefont {Y.}~\bibnamefont
  {Ashida}}, \bibinfo {author} {\bibfnamefont {Z.}~\bibnamefont {Gong}}, \ and\
  \bibinfo {author} {\bibfnamefont {M.}~\bibnamefont {Ueda}},\ }\href@noop {}
  {\bibfield  {journal} {\bibinfo  {journal} {arXiv preprint arXiv:2006.01837}\
  } (\bibinfo {year} {2020})}\BibitemShut {NoStop}%
\bibitem [{\citenamefont {Shen}\ \emph {et~al.}(2017)\citenamefont {Shen},
  \citenamefont {Zhen},\ and\ \citenamefont {Fu}}]{HShen2017_non-Hermi}%
  \BibitemOpen
  \bibfield  {author} {\bibinfo {author} {\bibfnamefont {H.}~\bibnamefont
  {Shen}}, \bibinfo {author} {\bibfnamefont {B.}~\bibnamefont {Zhen}}, \ and\
  \bibinfo {author} {\bibfnamefont {L.}~\bibnamefont {Fu}},\ }\href@noop {}
  {\bibfield  {journal} {\bibinfo  {journal} {arXiv preprint arXiv:1706.07435}\
  } (\bibinfo {year} {2017})}\BibitemShut {NoStop}%
\bibitem [{\citenamefont {Xu}\ \emph {et~al.}(2017)\citenamefont {Xu},
  \citenamefont {Wang},\ and\ \citenamefont
  {Duan}}]{YXuPRL17_exceptional_ring}%
  \BibitemOpen
  \bibfield  {author} {\bibinfo {author} {\bibfnamefont {Y.}~\bibnamefont
  {Xu}}, \bibinfo {author} {\bibfnamefont {S.-T.}\ \bibnamefont {Wang}}, \ and\
  \bibinfo {author} {\bibfnamefont {L.-M.}\ \bibnamefont {Duan}},\ }\href
  {\doibase 10.1103/PhysRevLett.118.045701} {\bibfield  {journal} {\bibinfo
  {journal} {Phys. Rev. Lett.}\ }\textbf {\bibinfo {volume} {118}},\ \bibinfo
  {pages} {045701} (\bibinfo {year} {2017})}\BibitemShut {NoStop}%
\bibitem [{\citenamefont {Budich}\ \emph {et~al.}(2019)\citenamefont {Budich},
  \citenamefont {Carlstr\"om}, \citenamefont {4Kunst},\ and\ \citenamefont
  {Bergholtz}}]{Budich_SPERs_PRB19}%
  \BibitemOpen
  \bibfield  {author} {\bibinfo {author} {\bibfnamefont {J.~C.}\ \bibnamefont
  {Budich}}, \bibinfo {author} {\bibfnamefont {J.}~\bibnamefont {Carlstr\"om}},
  \bibinfo {author} {\bibfnamefont {F.~K.}\ \bibnamefont {4Kunst}}, \ and\
  \bibinfo {author} {\bibfnamefont {E.~J.}\ \bibnamefont {Bergholtz}},\ }\href
  {\doibase 10.1103/PhysRevB.99.041406} {\bibfield  {journal} {\bibinfo
  {journal} {Phys. Rev. B}\ }\textbf {\bibinfo {volume} {99}},\ \bibinfo
  {pages} {041406} (\bibinfo {year} {2019})}\BibitemShut {NoStop}%
\bibitem [{\citenamefont {Okugawa}\ and\ \citenamefont
  {Yokoyama}(2019)}]{Okugawa_SPERs_PRB19}%
  \BibitemOpen
  \bibfield  {author} {\bibinfo {author} {\bibfnamefont {R.}~\bibnamefont
  {Okugawa}}\ and\ \bibinfo {author} {\bibfnamefont {T.}~\bibnamefont
  {Yokoyama}},\ }\href {\doibase 10.1103/PhysRevB.99.041202} {\bibfield
  {journal} {\bibinfo  {journal} {Phys. Rev. B}\ }\textbf {\bibinfo {volume}
  {99}},\ \bibinfo {pages} {041202} (\bibinfo {year} {2019})}\BibitemShut
  {NoStop}%
\bibitem [{\citenamefont {Yoshida}\ \emph
  {et~al.}(2019{\natexlab{a}})\citenamefont {Yoshida}, \citenamefont {Peters},
  \citenamefont {Kawakami},\ and\ \citenamefont
  {Hatsugai}}]{Yoshida_SPERs_PRB19}%
  \BibitemOpen
  \bibfield  {author} {\bibinfo {author} {\bibfnamefont {T.}~\bibnamefont
  {Yoshida}}, \bibinfo {author} {\bibfnamefont {R.}~\bibnamefont {Peters}},
  \bibinfo {author} {\bibfnamefont {N.}~\bibnamefont {Kawakami}}, \ and\
  \bibinfo {author} {\bibfnamefont {Y.}~\bibnamefont {Hatsugai}},\ }\href
  {\doibase 10.1103/PhysRevB.99.121101} {\bibfield  {journal} {\bibinfo
  {journal} {Phys. Rev. B}\ }\textbf {\bibinfo {volume} {99}},\ \bibinfo
  {pages} {121101} (\bibinfo {year} {2019}{\natexlab{a}})}\BibitemShut
  {NoStop}%
\bibitem [{\citenamefont {Zhou}\ \emph {et~al.}(2019)\citenamefont {Zhou},
  \citenamefont {Lee}, \citenamefont {Liu},\ and\ \citenamefont
  {Zhen}}]{Zhou_SPERs_Optica19}%
  \BibitemOpen
  \bibfield  {author} {\bibinfo {author} {\bibfnamefont {H.}~\bibnamefont
  {Zhou}}, \bibinfo {author} {\bibfnamefont {J.~Y.}\ \bibnamefont {Lee}},
  \bibinfo {author} {\bibfnamefont {S.}~\bibnamefont {Liu}}, \ and\ \bibinfo
  {author} {\bibfnamefont {B.}~\bibnamefont {Zhen}},\ }\href {\doibase
  10.1364/OPTICA.6.000190} {\bibfield  {journal} {\bibinfo  {journal} {Optica}\
  }\textbf {\bibinfo {volume} {6}},\ \bibinfo {pages} {190} (\bibinfo {year}
  {2019})}\BibitemShut {NoStop}%
\bibitem [{\citenamefont {Kawabata}\ \emph
  {et~al.}(2019{\natexlab{b}})\citenamefont {Kawabata}, \citenamefont
  {Bessho},\ and\ \citenamefont {Sato}}]{Kawabata_gapless_PRL19}%
  \BibitemOpen
  \bibfield  {author} {\bibinfo {author} {\bibfnamefont {K.}~\bibnamefont
  {Kawabata}}, \bibinfo {author} {\bibfnamefont {T.}~\bibnamefont {Bessho}}, \
  and\ \bibinfo {author} {\bibfnamefont {M.}~\bibnamefont {Sato}},\ }\href
  {\doibase 10.1103/PhysRevLett.123.066405} {\bibfield  {journal} {\bibinfo
  {journal} {Phys. Rev. Lett.}\ }\textbf {\bibinfo {volume} {123}},\ \bibinfo
  {pages} {066405} (\bibinfo {year} {2019}{\natexlab{b}})}\BibitemShut
  {NoStop}%
\bibitem [{\citenamefont {Martinez~Alvarez}\ \emph {et~al.}(2018)\citenamefont
  {Martinez~Alvarez}, \citenamefont {Barrios~Vargas},\ and\ \citenamefont
  {Foa~Torres}}]{Alvarez_nHSkin_PRB18}%
  \BibitemOpen
  \bibfield  {author} {\bibinfo {author} {\bibfnamefont {V.~M.}\ \bibnamefont
  {Martinez~Alvarez}}, \bibinfo {author} {\bibfnamefont {J.~E.}\ \bibnamefont
  {Barrios~Vargas}}, \ and\ \bibinfo {author} {\bibfnamefont {L.~E.~F.}\
  \bibnamefont {Foa~Torres}},\ }\href {\doibase 10.1103/PhysRevB.97.121401}
  {\bibfield  {journal} {\bibinfo  {journal} {Phys. Rev. B}\ }\textbf {\bibinfo
  {volume} {97}},\ \bibinfo {pages} {121401} (\bibinfo {year}
  {2018})}\BibitemShut {NoStop}%
\bibitem [{\citenamefont {Yao}\ and\ \citenamefont
  {Wang}(2018)}]{SYao_nHSkin-1D_PRL18}%
  \BibitemOpen
  \bibfield  {author} {\bibinfo {author} {\bibfnamefont {S.}~\bibnamefont
  {Yao}}\ and\ \bibinfo {author} {\bibfnamefont {Z.}~\bibnamefont {Wang}},\
  }\href {\doibase 10.1103/PhysRevLett.121.086803} {\bibfield  {journal}
  {\bibinfo  {journal} {Phys. Rev. Lett.}\ }\textbf {\bibinfo {volume} {121}},\
  \bibinfo {pages} {086803} (\bibinfo {year} {2018})}\BibitemShut {NoStop}%
\bibitem [{\citenamefont {Yao}\ \emph {et~al.}(2018)\citenamefont {Yao},
  \citenamefont {Song},\ and\ \citenamefont {Wang}}]{SYao_nHSkin-2D_PRL18}%
  \BibitemOpen
  \bibfield  {author} {\bibinfo {author} {\bibfnamefont {S.}~\bibnamefont
  {Yao}}, \bibinfo {author} {\bibfnamefont {F.}~\bibnamefont {Song}}, \ and\
  \bibinfo {author} {\bibfnamefont {Z.}~\bibnamefont {Wang}},\ }\href {\doibase
  10.1103/PhysRevLett.121.136802} {\bibfield  {journal} {\bibinfo  {journal}
  {Phys. Rev. Lett.}\ }\textbf {\bibinfo {volume} {121}},\ \bibinfo {pages}
  {136802} (\bibinfo {year} {2018})}\BibitemShut {NoStop}%
\bibitem [{\citenamefont {Lee}\ and\ \citenamefont
  {Thomale}(2019)}]{Lee_Skin19}%
  \BibitemOpen
  \bibfield  {author} {\bibinfo {author} {\bibfnamefont {C.~H.}\ \bibnamefont
  {Lee}}\ and\ \bibinfo {author} {\bibfnamefont {R.}~\bibnamefont {Thomale}},\
  }\href {\doibase 10.1103/PhysRevB.99.201103} {\bibfield  {journal} {\bibinfo
  {journal} {Phys. Rev. B}\ }\textbf {\bibinfo {volume} {99}},\ \bibinfo
  {pages} {201103} (\bibinfo {year} {2019})}\BibitemShut {NoStop}%
\bibitem [{\citenamefont {Zhang}\ \emph {et~al.}(2019)\citenamefont {Zhang},
  \citenamefont {Yang},\ and\ \citenamefont {Fang}}]{Zhang_BECskin19}%
  \BibitemOpen
  \bibfield  {author} {\bibinfo {author} {\bibfnamefont {K.}~\bibnamefont
  {Zhang}}, \bibinfo {author} {\bibfnamefont {Z.}~\bibnamefont {Yang}}, \ and\
  \bibinfo {author} {\bibfnamefont {C.}~\bibnamefont {Fang}},\ }\href@noop {}
  {\bibfield  {journal} {\bibinfo  {journal} {arXiv preprint arXiv:1910.01131}\
  } (\bibinfo {year} {2019})}\BibitemShut {NoStop}%
\bibitem [{\citenamefont {Okuma}\ \emph {et~al.}(2020)\citenamefont {Okuma},
  \citenamefont {Kawabata}, \citenamefont {Shiozaki},\ and\ \citenamefont
  {Sato}}]{Okuma_BECskin19}%
  \BibitemOpen
  \bibfield  {author} {\bibinfo {author} {\bibfnamefont {N.}~\bibnamefont
  {Okuma}}, \bibinfo {author} {\bibfnamefont {K.}~\bibnamefont {Kawabata}},
  \bibinfo {author} {\bibfnamefont {K.}~\bibnamefont {Shiozaki}}, \ and\
  \bibinfo {author} {\bibfnamefont {M.}~\bibnamefont {Sato}},\ }\href {\doibase
  10.1103/PhysRevLett.124.086801} {\bibfield  {journal} {\bibinfo  {journal}
  {Phys. Rev. Lett.}\ }\textbf {\bibinfo {volume} {124}},\ \bibinfo {pages}
  {086801} (\bibinfo {year} {2020})}\BibitemShut {NoStop}%
\bibitem [{\citenamefont {Okuma}\ and\ \citenamefont
  {Sato}(2020{\natexlab{a}})}]{Okuma_PS_PRB20}%
  \BibitemOpen
  \bibfield  {author} {\bibinfo {author} {\bibfnamefont {N.}~\bibnamefont
  {Okuma}}\ and\ \bibinfo {author} {\bibfnamefont {M.}~\bibnamefont {Sato}},\
  }\href {\doibase 10.1103/PhysRevB.102.014203} {\bibfield  {journal} {\bibinfo
   {journal} {Phys. Rev. B}\ }\textbf {\bibinfo {volume} {102}},\ \bibinfo
  {pages} {014203} (\bibinfo {year} {2020}{\natexlab{a}})}\BibitemShut
  {NoStop}%
\bibitem [{\citenamefont {Mu}\ \emph {et~al.}(2020)\citenamefont {Mu},
  \citenamefont {Lee}, \citenamefont {Li},\ and\ \citenamefont
  {Gong}}]{Mu_MbdySkin_PRB20}%
  \BibitemOpen
  \bibfield  {author} {\bibinfo {author} {\bibfnamefont {S.}~\bibnamefont
  {Mu}}, \bibinfo {author} {\bibfnamefont {C.~H.}\ \bibnamefont {Lee}},
  \bibinfo {author} {\bibfnamefont {L.}~\bibnamefont {Li}}, \ and\ \bibinfo
  {author} {\bibfnamefont {J.}~\bibnamefont {Gong}},\ }\href {\doibase
  10.1103/PhysRevB.102.081115} {\bibfield  {journal} {\bibinfo  {journal}
  {Phys. Rev. B}\ }\textbf {\bibinfo {volume} {102}},\ \bibinfo {pages}
  {081115} (\bibinfo {year} {2020})}\BibitemShut {NoStop}%
\bibitem [{\citenamefont {Lee}(2020)}]{Lee_MbdySkin_arXiv20}%
  \BibitemOpen
  \bibfield  {author} {\bibinfo {author} {\bibfnamefont {C.~H.}\ \bibnamefont
  {Lee}},\ }\href@noop {} {\bibfield  {journal} {\bibinfo  {journal} {arXiv
  preprint arXiv:2006.01182}\ } (\bibinfo {year} {2020})}\BibitemShut {NoStop}%
\bibitem [{\citenamefont {Okugawa}\ \emph {et~al.}(2020)\citenamefont
  {Okugawa}, \citenamefont {Takahashi},\ and\ \citenamefont
  {Yokomizo}}]{Okugawa_HOSkin_arXiv2020}%
  \BibitemOpen
  \bibfield  {author} {\bibinfo {author} {\bibfnamefont {R.}~\bibnamefont
  {Okugawa}}, \bibinfo {author} {\bibfnamefont {R.}~\bibnamefont {Takahashi}},
  \ and\ \bibinfo {author} {\bibfnamefont {K.}~\bibnamefont {Yokomizo}},\
  }\href@noop {} {\bibfield  {journal} {\bibinfo  {journal} {arXiv preprint
  arXiv:2008.03721}\ } (\bibinfo {year} {2020})}\BibitemShut {NoStop}%
\bibitem [{\citenamefont {Kawabata}\ \emph {et~al.}(2020)\citenamefont
  {Kawabata}, \citenamefont {Sato},\ and\ \citenamefont
  {Shiozaki}}]{Kawabata_HOSkin_arXiv2020}%
  \BibitemOpen
  \bibfield  {author} {\bibinfo {author} {\bibfnamefont {K.}~\bibnamefont
  {Kawabata}}, \bibinfo {author} {\bibfnamefont {M.}~\bibnamefont {Sato}}, \
  and\ \bibinfo {author} {\bibfnamefont {K.}~\bibnamefont {Shiozaki}},\
  }\href@noop {} {\bibfield  {journal} {\bibinfo  {journal} {arXiv preprint
  arXiv:2008.07237}\ } (\bibinfo {year} {2020})}\BibitemShut {NoStop}%
\bibitem [{\citenamefont {Fu}\ and\ \citenamefont
  {Wan}(2020)}]{Fu_HOSkin_arXiv2020}%
  \BibitemOpen
  \bibfield  {author} {\bibinfo {author} {\bibfnamefont {Y.}~\bibnamefont
  {Fu}}\ and\ \bibinfo {author} {\bibfnamefont {S.}~\bibnamefont {Wan}},\
  }\href@noop {} {\bibfield  {journal} {\bibinfo  {journal} {arXiv preprint
  arXiv:2008.09033}\ } (\bibinfo {year} {2020})}\BibitemShut {NoStop}%
\bibitem [{\citenamefont {R{\"u}ter}\ \emph {et~al.}(2010)\citenamefont
  {R{\"u}ter}, \citenamefont {Makris}, \citenamefont {El-Ganainy},
  \citenamefont {Christodoulides}, \citenamefont {Segev},\ and\ \citenamefont
  {Kip}}]{Ruter_nHExp_NatPhys10}%
  \BibitemOpen
  \bibfield  {author} {\bibinfo {author} {\bibfnamefont {C.~E.}\ \bibnamefont
  {R{\"u}ter}}, \bibinfo {author} {\bibfnamefont {K.~G.}\ \bibnamefont
  {Makris}}, \bibinfo {author} {\bibfnamefont {R.}~\bibnamefont {El-Ganainy}},
  \bibinfo {author} {\bibfnamefont {D.~N.}\ \bibnamefont {Christodoulides}},
  \bibinfo {author} {\bibfnamefont {M.}~\bibnamefont {Segev}}, \ and\ \bibinfo
  {author} {\bibfnamefont {D.}~\bibnamefont {Kip}},\ }\href@noop {} {\bibfield
  {journal} {\bibinfo  {journal} {Nature physics}\ }\textbf {\bibinfo {volume}
  {6}},\ \bibinfo {pages} {192} (\bibinfo {year} {2010})}\BibitemShut {NoStop}%
\bibitem [{\citenamefont {Regensburger}\ \emph {et~al.}(2012)\citenamefont
  {Regensburger}, \citenamefont {Bersch}, \citenamefont {Miri}, \citenamefont
  {Onishchukov}, \citenamefont {Christodoulides},\ and\ \citenamefont
  {Peschel}}]{Regensburger_nHExp_Nat12}%
  \BibitemOpen
  \bibfield  {author} {\bibinfo {author} {\bibfnamefont {A.}~\bibnamefont
  {Regensburger}}, \bibinfo {author} {\bibfnamefont {C.}~\bibnamefont
  {Bersch}}, \bibinfo {author} {\bibfnamefont {M.-A.}\ \bibnamefont {Miri}},
  \bibinfo {author} {\bibfnamefont {G.}~\bibnamefont {Onishchukov}}, \bibinfo
  {author} {\bibfnamefont {D.~N.}\ \bibnamefont {Christodoulides}}, \ and\
  \bibinfo {author} {\bibfnamefont {U.}~\bibnamefont {Peschel}},\ }\href@noop
  {} {\bibfield  {journal} {\bibinfo  {journal} {Nature}\ }\textbf {\bibinfo
  {volume} {488}},\ \bibinfo {pages} {167} (\bibinfo {year}
  {2012})}\BibitemShut {NoStop}%
\bibitem [{\citenamefont {Zhen}\ \emph {et~al.}(2015)\citenamefont {Zhen},
  \citenamefont {Hsu}, \citenamefont {Igarashi}, \citenamefont {Lu},
  \citenamefont {Kaminer}, \citenamefont {Pick}, \citenamefont {Chua},
  \citenamefont {Joannopoulos},\ and\ \citenamefont
  {Soljacic}}]{Zhen_AcciEP_Nat15}%
  \BibitemOpen
  \bibfield  {author} {\bibinfo {author} {\bibfnamefont {B.}~\bibnamefont
  {Zhen}}, \bibinfo {author} {\bibfnamefont {C.~W.}\ \bibnamefont {Hsu}},
  \bibinfo {author} {\bibfnamefont {Y.}~\bibnamefont {Igarashi}}, \bibinfo
  {author} {\bibfnamefont {L.}~\bibnamefont {Lu}}, \bibinfo {author}
  {\bibfnamefont {I.}~\bibnamefont {Kaminer}}, \bibinfo {author} {\bibfnamefont
  {A.}~\bibnamefont {Pick}}, \bibinfo {author} {\bibfnamefont {S.-L.}\
  \bibnamefont {Chua}}, \bibinfo {author} {\bibfnamefont {J.~D.}\ \bibnamefont
  {Joannopoulos}}, \ and\ \bibinfo {author} {\bibfnamefont {M.}~\bibnamefont
  {Soljacic}},\ }\href {http://dx.doi.org/10.1038/nature14889} {\bibfield
  {journal} {\bibinfo  {journal} {Nature}\ }\textbf {\bibinfo {volume} {525}},\
  \bibinfo {pages} {354 EP } (\bibinfo {year} {2015})}\BibitemShut {NoStop}%
\bibitem [{\citenamefont {Hassan}\ \emph {et~al.}(2017)\citenamefont {Hassan},
  \citenamefont {Zhen}, \citenamefont {Solja\ifmmode \check{c}\else
  \v{c}\fi{}i\ifmmode~\acute{c}\else \'{c}\fi{}}, \citenamefont {Khajavikhan},\
  and\ \citenamefont {Christodoulides}}]{Hassan_EP_PRL17}%
  \BibitemOpen
  \bibfield  {author} {\bibinfo {author} {\bibfnamefont {A.~U.}\ \bibnamefont
  {Hassan}}, \bibinfo {author} {\bibfnamefont {B.}~\bibnamefont {Zhen}},
  \bibinfo {author} {\bibfnamefont {M.}~\bibnamefont {Solja\ifmmode
  \check{c}\else \v{c}\fi{}i\ifmmode~\acute{c}\else \'{c}\fi{}}}, \bibinfo
  {author} {\bibfnamefont {M.}~\bibnamefont {Khajavikhan}}, \ and\ \bibinfo
  {author} {\bibfnamefont {D.~N.}\ \bibnamefont {Christodoulides}},\ }\href
  {\doibase 10.1103/PhysRevLett.118.093002} {\bibfield  {journal} {\bibinfo
  {journal} {Phys. Rev. Lett.}\ }\textbf {\bibinfo {volume} {118}},\ \bibinfo
  {pages} {093002} (\bibinfo {year} {2017})}\BibitemShut {NoStop}%
\bibitem [{\citenamefont {Zhou}\ \emph {et~al.}(2018)\citenamefont {Zhou},
  \citenamefont {Peng}, \citenamefont {Yoon}, \citenamefont {Hsu},
  \citenamefont {Nelson}, \citenamefont {Fu}, \citenamefont {Joannopoulos},
  \citenamefont {Solja{\v c}i{\'c}},\ and\ \citenamefont
  {Zhen}}]{Zhou_FermiArcPH_Science18}%
  \BibitemOpen
  \bibfield  {author} {\bibinfo {author} {\bibfnamefont {H.}~\bibnamefont
  {Zhou}}, \bibinfo {author} {\bibfnamefont {C.}~\bibnamefont {Peng}}, \bibinfo
  {author} {\bibfnamefont {Y.}~\bibnamefont {Yoon}}, \bibinfo {author}
  {\bibfnamefont {C.~W.}\ \bibnamefont {Hsu}}, \bibinfo {author} {\bibfnamefont
  {K.~A.}\ \bibnamefont {Nelson}}, \bibinfo {author} {\bibfnamefont
  {L.}~\bibnamefont {Fu}}, \bibinfo {author} {\bibfnamefont {J.~D.}\
  \bibnamefont {Joannopoulos}}, \bibinfo {author} {\bibfnamefont
  {M.}~\bibnamefont {Solja{\v c}i{\'c}}}, \ and\ \bibinfo {author}
  {\bibfnamefont {B.}~\bibnamefont {Zhen}},\ }\href {\doibase
  10.1126/science.aap9859} {\bibfield  {journal} {\bibinfo  {journal}
  {Science}\ }\textbf {\bibinfo {volume} {359}},\ \bibinfo {pages} {1009}
  (\bibinfo {year} {2018})},\ \Eprint
  {http://arxiv.org/abs/https://science.sciencemag.org/content/359/6379/1009.full.pdf}
  {https://science.sciencemag.org/content/359/6379/1009.full.pdf} \BibitemShut
  {NoStop}%
\bibitem [{\citenamefont {Takata}\ and\ \citenamefont
  {Notomi}(2018)}]{Takata_pSSH_PRL18}%
  \BibitemOpen
  \bibfield  {author} {\bibinfo {author} {\bibfnamefont {K.}~\bibnamefont
  {Takata}}\ and\ \bibinfo {author} {\bibfnamefont {M.}~\bibnamefont
  {Notomi}},\ }\href {\doibase 10.1103/PhysRevLett.121.213902} {\bibfield
  {journal} {\bibinfo  {journal} {Phys. Rev. Lett.}\ }\textbf {\bibinfo
  {volume} {121}},\ \bibinfo {pages} {213902} (\bibinfo {year}
  {2018})}\BibitemShut {NoStop}%
\bibitem [{\citenamefont {Ozawa}\ \emph {et~al.}(2019)\citenamefont {Ozawa},
  \citenamefont {Price}, \citenamefont {Amo}, \citenamefont {Goldman},
  \citenamefont {Hafezi}, \citenamefont {Lu}, \citenamefont {Rechtsman},
  \citenamefont {Schuster}, \citenamefont {Simon}, \citenamefont {Zilberberg},\
  and\ \citenamefont {Carusotto}}]{Ozawa_TopoPhoto_RMP19}%
  \BibitemOpen
  \bibfield  {author} {\bibinfo {author} {\bibfnamefont {T.}~\bibnamefont
  {Ozawa}}, \bibinfo {author} {\bibfnamefont {H.~M.}\ \bibnamefont {Price}},
  \bibinfo {author} {\bibfnamefont {A.}~\bibnamefont {Amo}}, \bibinfo {author}
  {\bibfnamefont {N.}~\bibnamefont {Goldman}}, \bibinfo {author} {\bibfnamefont
  {M.}~\bibnamefont {Hafezi}}, \bibinfo {author} {\bibfnamefont
  {L.}~\bibnamefont {Lu}}, \bibinfo {author} {\bibfnamefont {M.~C.}\
  \bibnamefont {Rechtsman}}, \bibinfo {author} {\bibfnamefont {D.}~\bibnamefont
  {Schuster}}, \bibinfo {author} {\bibfnamefont {J.}~\bibnamefont {Simon}},
  \bibinfo {author} {\bibfnamefont {O.}~\bibnamefont {Zilberberg}}, \ and\
  \bibinfo {author} {\bibfnamefont {I.}~\bibnamefont {Carusotto}},\ }\href
  {\doibase 10.1103/RevModPhys.91.015006} {\bibfield  {journal} {\bibinfo
  {journal} {Rev. Mod. Phys.}\ }\textbf {\bibinfo {volume} {91}},\ \bibinfo
  {pages} {015006} (\bibinfo {year} {2019})}\BibitemShut {NoStop}%
\bibitem [{\citenamefont {Xiao}\ \emph {et~al.}(2020)\citenamefont {Xiao},
  \citenamefont {Deng}, \citenamefont {Wang}, \citenamefont {Zhu},
  \citenamefont {Wang}, \citenamefont {Yi},\ and\ \citenamefont
  {Xue}}]{Xiao_nHSkin_Exp_NatPhys19}%
  \BibitemOpen
  \bibfield  {author} {\bibinfo {author} {\bibfnamefont {L.}~\bibnamefont
  {Xiao}}, \bibinfo {author} {\bibfnamefont {T.}~\bibnamefont {Deng}}, \bibinfo
  {author} {\bibfnamefont {K.}~\bibnamefont {Wang}}, \bibinfo {author}
  {\bibfnamefont {G.}~\bibnamefont {Zhu}}, \bibinfo {author} {\bibfnamefont
  {Z.}~\bibnamefont {Wang}}, \bibinfo {author} {\bibfnamefont {W.}~\bibnamefont
  {Yi}}, \ and\ \bibinfo {author} {\bibfnamefont {P.}~\bibnamefont {Xue}},\
  }\href {\doibase 10.1038/s41567-020-0836-6} {\bibfield  {journal} {\bibinfo
  {journal} {Nature Physics}\ }\textbf {\bibinfo {volume} {16}},\ \bibinfo
  {pages} {761} (\bibinfo {year} {2020})}\BibitemShut {NoStop}%
\bibitem [{\citenamefont {Diehl}\ \emph {et~al.}(2011)\citenamefont {Diehl},
  \citenamefont {Rico}, \citenamefont {Baranov},\ and\ \citenamefont
  {Zoller}}]{Diehl_DissCher_NatPhys11}%
  \BibitemOpen
  \bibfield  {author} {\bibinfo {author} {\bibfnamefont {S.}~\bibnamefont
  {Diehl}}, \bibinfo {author} {\bibfnamefont {E.}~\bibnamefont {Rico}},
  \bibinfo {author} {\bibfnamefont {M.~A.}\ \bibnamefont {Baranov}}, \ and\
  \bibinfo {author} {\bibfnamefont {P.}~\bibnamefont {Zoller}},\ }\href
  {\doibase 10.1038/nphys2106} {\bibfield  {journal} {\bibinfo  {journal}
  {Nature Physics}\ }\textbf {\bibinfo {volume} {7}},\ \bibinfo {pages} {971}
  (\bibinfo {year} {2011})}\BibitemShut {NoStop}%
\bibitem [{\citenamefont {Bardyn}\ \emph {et~al.}(2013)\citenamefont {Bardyn},
  \citenamefont {Baranov}, \citenamefont {Kraus}, \citenamefont {Rico},
  \citenamefont {{\.{I}}mamo{\u{g}}lu}, \citenamefont {Zoller},\ and\
  \citenamefont {Diehl}}]{Bardyn_DissCher_NJP2013}%
  \BibitemOpen
  \bibfield  {author} {\bibinfo {author} {\bibfnamefont {C.-E.}\ \bibnamefont
  {Bardyn}}, \bibinfo {author} {\bibfnamefont {M.~A.}\ \bibnamefont {Baranov}},
  \bibinfo {author} {\bibfnamefont {C.~V.}\ \bibnamefont {Kraus}}, \bibinfo
  {author} {\bibfnamefont {E.}~\bibnamefont {Rico}}, \bibinfo {author}
  {\bibfnamefont {A.}~\bibnamefont {{\.{I}}mamo{\u{g}}lu}}, \bibinfo {author}
  {\bibfnamefont {P.}~\bibnamefont {Zoller}}, \ and\ \bibinfo {author}
  {\bibfnamefont {S.}~\bibnamefont {Diehl}},\ }\href {\doibase
  10.1088/1367-2630/15/8/085001} {\bibfield  {journal} {\bibinfo  {journal}
  {New Journal of Physics}\ }\textbf {\bibinfo {volume} {15}},\ \bibinfo
  {pages} {085001} (\bibinfo {year} {2013})}\BibitemShut {NoStop}%
\bibitem [{\citenamefont {Rivas}\ \emph {et~al.}(2013)\citenamefont {Rivas},
  \citenamefont {Viyuela},\ and\ \citenamefont
  {Martin-Delgado}}]{Rivas_DissCher_PRB13}%
  \BibitemOpen
  \bibfield  {author} {\bibinfo {author} {\bibfnamefont {A.}~\bibnamefont
  {Rivas}}, \bibinfo {author} {\bibfnamefont {O.}~\bibnamefont {Viyuela}}, \
  and\ \bibinfo {author} {\bibfnamefont {M.~A.}\ \bibnamefont
  {Martin-Delgado}},\ }\href {\doibase 10.1103/PhysRevB.88.155141} {\bibfield
  {journal} {\bibinfo  {journal} {Phys. Rev. B}\ }\textbf {\bibinfo {volume}
  {88}},\ \bibinfo {pages} {155141} (\bibinfo {year} {2013})}\BibitemShut
  {NoStop}%
\bibitem [{\citenamefont {Budich}\ \emph {et~al.}(2015)\citenamefont {Budich},
  \citenamefont {Zoller},\ and\ \citenamefont {Diehl}}]{Budich_DissCher_PRA15}%
  \BibitemOpen
  \bibfield  {author} {\bibinfo {author} {\bibfnamefont {J.~C.}\ \bibnamefont
  {Budich}}, \bibinfo {author} {\bibfnamefont {P.}~\bibnamefont {Zoller}}, \
  and\ \bibinfo {author} {\bibfnamefont {S.}~\bibnamefont {Diehl}},\ }\href
  {\doibase 10.1103/PhysRevA.91.042117} {\bibfield  {journal} {\bibinfo
  {journal} {Phys. Rev. A}\ }\textbf {\bibinfo {volume} {91}},\ \bibinfo
  {pages} {042117} (\bibinfo {year} {2015})}\BibitemShut {NoStop}%
\bibitem [{\citenamefont {Budich}\ and\ \citenamefont
  {Diehl}(2015)}]{Budich_DissCher_PRB15}%
  \BibitemOpen
  \bibfield  {author} {\bibinfo {author} {\bibfnamefont {J.~C.}\ \bibnamefont
  {Budich}}\ and\ \bibinfo {author} {\bibfnamefont {S.}~\bibnamefont {Diehl}},\
  }\href {\doibase 10.1103/PhysRevB.91.165140} {\bibfield  {journal} {\bibinfo
  {journal} {Phys. Rev. B}\ }\textbf {\bibinfo {volume} {91}},\ \bibinfo
  {pages} {165140} (\bibinfo {year} {2015})}\BibitemShut {NoStop}%
\bibitem [{\citenamefont {Gong}\ \emph {et~al.}(2017)\citenamefont {Gong},
  \citenamefont {Higashikawa},\ and\ \citenamefont {Ueda}}]{ZPGong_PRL17}%
  \BibitemOpen
  \bibfield  {author} {\bibinfo {author} {\bibfnamefont {Z.}~\bibnamefont
  {Gong}}, \bibinfo {author} {\bibfnamefont {S.}~\bibnamefont {Higashikawa}}, \
  and\ \bibinfo {author} {\bibfnamefont {M.}~\bibnamefont {Ueda}},\ }\href
  {\doibase 10.1103/PhysRevLett.118.200401} {\bibfield  {journal} {\bibinfo
  {journal} {Phys. Rev. Lett.}\ }\textbf {\bibinfo {volume} {118}},\ \bibinfo
  {pages} {200401} (\bibinfo {year} {2017})}\BibitemShut {NoStop}%
\bibitem [{\citenamefont {Lieu}\ \emph {et~al.}(2020)\citenamefont {Lieu},
  \citenamefont {McGinley},\ and\ \citenamefont
  {Cooper}}]{Lieu_Liouclass_PRL20}%
  \BibitemOpen
  \bibfield  {author} {\bibinfo {author} {\bibfnamefont {S.}~\bibnamefont
  {Lieu}}, \bibinfo {author} {\bibfnamefont {M.}~\bibnamefont {McGinley}}, \
  and\ \bibinfo {author} {\bibfnamefont {N.~R.}\ \bibnamefont {Cooper}},\
  }\href {\doibase 10.1103/PhysRevLett.124.040401} {\bibfield  {journal}
  {\bibinfo  {journal} {Phys. Rev. Lett.}\ }\textbf {\bibinfo {volume} {124}},\
  \bibinfo {pages} {040401} (\bibinfo {year} {2020})}\BibitemShut {NoStop}%
\bibitem [{\citenamefont {Yoshida}\ \emph
  {et~al.}(2019{\natexlab{b}})\citenamefont {Yoshida}, \citenamefont {Kudo},\
  and\ \citenamefont {Hatsugai}}]{Yoshida_nHFQH19}%
  \BibitemOpen
  \bibfield  {author} {\bibinfo {author} {\bibfnamefont {T.}~\bibnamefont
  {Yoshida}}, \bibinfo {author} {\bibfnamefont {K.}~\bibnamefont {Kudo}}, \
  and\ \bibinfo {author} {\bibfnamefont {Y.}~\bibnamefont {Hatsugai}},\ }\href
  {\doibase 10.1038/s41598-019-53253-8} {\bibfield  {journal} {\bibinfo
  {journal} {Scientific Reports}\ }\textbf {\bibinfo {volume} {9}},\ \bibinfo
  {pages} {16895} (\bibinfo {year} {2019}{\natexlab{b}})}\BibitemShut {NoStop}%
\bibitem [{\citenamefont {Yoshida}\ \emph
  {et~al.}(2020{\natexlab{b}})\citenamefont {Yoshida}, \citenamefont {Kudo},
  \citenamefont {Katsura},\ and\ \citenamefont
  {Hatsugai}}]{Yoshida_nHFQHJ_PRR20}%
  \BibitemOpen
  \bibfield  {author} {\bibinfo {author} {\bibfnamefont {T.}~\bibnamefont
  {Yoshida}}, \bibinfo {author} {\bibfnamefont {K.}~\bibnamefont {Kudo}},
  \bibinfo {author} {\bibfnamefont {H.}~\bibnamefont {Katsura}}, \ and\
  \bibinfo {author} {\bibfnamefont {Y.}~\bibnamefont {Hatsugai}},\ }\href
  {\doibase 10.1103/PhysRevResearch.2.033428} {\bibfield  {journal} {\bibinfo
  {journal} {Phys. Rev. Research}\ }\textbf {\bibinfo {volume} {2}},\ \bibinfo
  {pages} {033428} (\bibinfo {year} {2020}{\natexlab{b}})}\BibitemShut
  {NoStop}%
\bibitem [{\citenamefont {Yoshida}\ and\ \citenamefont
  {Hatsugai}(2019)}]{Yoshida_SPERs_mech19}%
  \BibitemOpen
  \bibfield  {author} {\bibinfo {author} {\bibfnamefont {T.}~\bibnamefont
  {Yoshida}}\ and\ \bibinfo {author} {\bibfnamefont {Y.}~\bibnamefont
  {Hatsugai}},\ }\href {\doibase 10.1103/PhysRevB.100.054109} {\bibfield
  {journal} {\bibinfo  {journal} {Phys. Rev. B}\ }\textbf {\bibinfo {volume}
  {100}},\ \bibinfo {pages} {054109} (\bibinfo {year} {2019})}\BibitemShut
  {NoStop}%
\bibitem [{\citenamefont {Ghatak}\ \emph {et~al.}(2019)\citenamefont {Ghatak},
  \citenamefont {Brandenbourger}, \citenamefont {van Wezel},\ and\
  \citenamefont {Coulais}}]{Ghatak_Mech_nHskin_arXiv19}%
  \BibitemOpen
  \bibfield  {author} {\bibinfo {author} {\bibfnamefont {A.}~\bibnamefont
  {Ghatak}}, \bibinfo {author} {\bibfnamefont {M.}~\bibnamefont
  {Brandenbourger}}, \bibinfo {author} {\bibfnamefont {J.}~\bibnamefont {van
  Wezel}}, \ and\ \bibinfo {author} {\bibfnamefont {C.}~\bibnamefont
  {Coulais}},\ }\href@noop {} {\bibfield  {journal} {\bibinfo  {journal} {arXiv
  preprint arXiv:1907.11619}\ } (\bibinfo {year} {2019})}\BibitemShut {NoStop}%
\bibitem [{\citenamefont {Scheibner}\ \emph {et~al.}(2020)\citenamefont
  {Scheibner}, \citenamefont {Irvine},\ and\ \citenamefont
  {Vitelli}}]{Scheibner_nHmech_PRL20}%
  \BibitemOpen
  \bibfield  {author} {\bibinfo {author} {\bibfnamefont {C.}~\bibnamefont
  {Scheibner}}, \bibinfo {author} {\bibfnamefont {W.~T.~M.}\ \bibnamefont
  {Irvine}}, \ and\ \bibinfo {author} {\bibfnamefont {V.}~\bibnamefont
  {Vitelli}},\ }\href {\doibase 10.1103/PhysRevLett.125.118001} {\bibfield
  {journal} {\bibinfo  {journal} {Phys. Rev. Lett.}\ }\textbf {\bibinfo
  {volume} {125}},\ \bibinfo {pages} {118001} (\bibinfo {year}
  {2020})}\BibitemShut {NoStop}%
\bibitem [{\citenamefont {Helbig}\ \emph {et~al.}(2019)\citenamefont {Helbig},
  \citenamefont {Hofmann}, \citenamefont {Imhof}, \citenamefont {Abdelghany},
  \citenamefont {Kiessling}, \citenamefont {Molenkamp}, \citenamefont {Lee},
  \citenamefont {Szameit}, \citenamefont {Greiter},\ and\ \citenamefont
  {Thomale}}]{Helbig_ExpSkin_19}%
  \BibitemOpen
  \bibfield  {author} {\bibinfo {author} {\bibfnamefont {T.}~\bibnamefont
  {Helbig}}, \bibinfo {author} {\bibfnamefont {T.}~\bibnamefont {Hofmann}},
  \bibinfo {author} {\bibfnamefont {S.}~\bibnamefont {Imhof}}, \bibinfo
  {author} {\bibfnamefont {M.}~\bibnamefont {Abdelghany}}, \bibinfo {author}
  {\bibfnamefont {T.}~\bibnamefont {Kiessling}}, \bibinfo {author}
  {\bibfnamefont {L.~W.}\ \bibnamefont {Molenkamp}}, \bibinfo {author}
  {\bibfnamefont {C.~H.}\ \bibnamefont {Lee}}, \bibinfo {author} {\bibfnamefont
  {A.}~\bibnamefont {Szameit}}, \bibinfo {author} {\bibfnamefont
  {M.}~\bibnamefont {Greiter}}, \ and\ \bibinfo {author} {\bibfnamefont
  {R.}~\bibnamefont {Thomale}},\ }\href@noop {} {\bibfield  {journal} {\bibinfo
   {journal} {arXiv preprint arXiv:1907.11562}\ } (\bibinfo {year}
  {2019})}\BibitemShut {NoStop}%
\bibitem [{\citenamefont {Hofmann}\ \emph {et~al.}(2019)\citenamefont
  {Hofmann}, \citenamefont {Helbig}, \citenamefont {Schindler}, \citenamefont
  {Salgo}, \citenamefont {Brzezi{\'n}ska}, \citenamefont {Greiter},
  \citenamefont {Kiessling}, \citenamefont {Wolf}, \citenamefont {Vollhardt},
  \citenamefont {Kaba{\v{s}}i} \emph {et~al.}}]{Hofmann_ExpRecipSkin_19}%
  \BibitemOpen
  \bibfield  {author} {\bibinfo {author} {\bibfnamefont {T.}~\bibnamefont
  {Hofmann}}, \bibinfo {author} {\bibfnamefont {T.}~\bibnamefont {Helbig}},
  \bibinfo {author} {\bibfnamefont {F.}~\bibnamefont {Schindler}}, \bibinfo
  {author} {\bibfnamefont {N.}~\bibnamefont {Salgo}}, \bibinfo {author}
  {\bibfnamefont {M.}~\bibnamefont {Brzezi{\'n}ska}}, \bibinfo {author}
  {\bibfnamefont {M.}~\bibnamefont {Greiter}}, \bibinfo {author} {\bibfnamefont
  {T.}~\bibnamefont {Kiessling}}, \bibinfo {author} {\bibfnamefont
  {D.}~\bibnamefont {Wolf}}, \bibinfo {author} {\bibfnamefont {A.}~\bibnamefont
  {Vollhardt}}, \bibinfo {author} {\bibfnamefont {A.}~\bibnamefont
  {Kaba{\v{s}}i}},  \emph {et~al.},\ }\href@noop {} {\bibfield  {journal}
  {\bibinfo  {journal} {arXiv preprint arXiv:1908.02759}\ } (\bibinfo {year}
  {2019})}\BibitemShut {NoStop}%
\bibitem [{\citenamefont {Yoshida}\ \emph
  {et~al.}(2020{\natexlab{c}})\citenamefont {Yoshida}, \citenamefont
  {Mizoguchi},\ and\ \citenamefont {Hatsugai}}]{Yoshida_MSkinPRR20}%
  \BibitemOpen
  \bibfield  {author} {\bibinfo {author} {\bibfnamefont {T.}~\bibnamefont
  {Yoshida}}, \bibinfo {author} {\bibfnamefont {T.}~\bibnamefont {Mizoguchi}},
  \ and\ \bibinfo {author} {\bibfnamefont {Y.}~\bibnamefont {Hatsugai}},\
  }\href {\doibase 10.1103/PhysRevResearch.2.022062} {\bibfield  {journal}
  {\bibinfo  {journal} {Phys. Rev. Research}\ }\textbf {\bibinfo {volume}
  {2}},\ \bibinfo {pages} {022062} (\bibinfo {year}
  {2020}{\natexlab{c}})}\BibitemShut {NoStop}%
\bibitem [{\citenamefont {Kozii}\ and\ \citenamefont
  {Fu}(2017)}]{VKozii_nH_arXiv17}%
  \BibitemOpen
  \bibfield  {author} {\bibinfo {author} {\bibfnamefont {V.}~\bibnamefont
  {Kozii}}\ and\ \bibinfo {author} {\bibfnamefont {L.}~\bibnamefont {Fu}},\
  }\href@noop {} {\bibfield  {journal} {\bibinfo  {journal} {arXiv preprint
  arXiv:1708.05841}\ } (\bibinfo {year} {2017})}\BibitemShut {NoStop}%
\bibitem [{\citenamefont {Yoshida}\ \emph {et~al.}(2018)\citenamefont
  {Yoshida}, \citenamefont {Peters},\ and\ \citenamefont
  {Kawakami}}]{Yoshida_EP_DMFT_PRB18}%
  \BibitemOpen
  \bibfield  {author} {\bibinfo {author} {\bibfnamefont {T.}~\bibnamefont
  {Yoshida}}, \bibinfo {author} {\bibfnamefont {R.}~\bibnamefont {Peters}}, \
  and\ \bibinfo {author} {\bibfnamefont {N.}~\bibnamefont {Kawakami}},\ }\href
  {\doibase 10.1103/PhysRevB.98.035141} {\bibfield  {journal} {\bibinfo
  {journal} {Phys. Rev. B}\ }\textbf {\bibinfo {volume} {98}},\ \bibinfo
  {pages} {035141} (\bibinfo {year} {2018})}\BibitemShut {NoStop}%
\bibitem [{\citenamefont {Shen}\ and\ \citenamefont
  {Fu}(2018)}]{HShen2018quantum_osci}%
  \BibitemOpen
  \bibfield  {author} {\bibinfo {author} {\bibfnamefont {H.}~\bibnamefont
  {Shen}}\ and\ \bibinfo {author} {\bibfnamefont {L.}~\bibnamefont {Fu}},\
  }\href@noop {} {\bibfield  {journal} {\bibinfo  {journal} {arXiv preprint
  arXiv:1802.03023}\ } (\bibinfo {year} {2018})}\BibitemShut {NoStop}%
\bibitem [{\citenamefont {Zyuzin}\ and\ \citenamefont
  {Zyuzin}(2018)}]{Zyuzin_nHEP_PRB18}%
  \BibitemOpen
  \bibfield  {author} {\bibinfo {author} {\bibfnamefont {A.~A.}\ \bibnamefont
  {Zyuzin}}\ and\ \bibinfo {author} {\bibfnamefont {A.~Y.}\ \bibnamefont
  {Zyuzin}},\ }\href {\doibase 10.1103/PhysRevB.97.041203} {\bibfield
  {journal} {\bibinfo  {journal} {Phys. Rev. B}\ }\textbf {\bibinfo {volume}
  {97}},\ \bibinfo {pages} {041203} (\bibinfo {year} {2018})}\BibitemShut
  {NoStop}%
\bibitem [{\citenamefont {Papaj}\ \emph {et~al.}(2019)\citenamefont {Papaj},
  \citenamefont {Isobe},\ and\ \citenamefont {Fu}}]{Papaji_nHEP_PRB19}%
  \BibitemOpen
  \bibfield  {author} {\bibinfo {author} {\bibfnamefont {M.}~\bibnamefont
  {Papaj}}, \bibinfo {author} {\bibfnamefont {H.}~\bibnamefont {Isobe}}, \ and\
  \bibinfo {author} {\bibfnamefont {L.}~\bibnamefont {Fu}},\ }\href {\doibase
  10.1103/PhysRevB.99.201107} {\bibfield  {journal} {\bibinfo  {journal} {Phys.
  Rev. B}\ }\textbf {\bibinfo {volume} {99}},\ \bibinfo {pages} {201107}
  (\bibinfo {year} {2019})}\BibitemShut {NoStop}%
\bibitem [{\citenamefont {Kimura}\ \emph {et~al.}(2019)\citenamefont {Kimura},
  \citenamefont {Yoshida},\ and\ \citenamefont
  {Kawakami}}]{Kimura_SPERs_PRB19}%
  \BibitemOpen
  \bibfield  {author} {\bibinfo {author} {\bibfnamefont {K.}~\bibnamefont
  {Kimura}}, \bibinfo {author} {\bibfnamefont {T.}~\bibnamefont {Yoshida}}, \
  and\ \bibinfo {author} {\bibfnamefont {N.}~\bibnamefont {Kawakami}},\ }\href
  {\doibase 10.1103/PhysRevB.100.115124} {\bibfield  {journal} {\bibinfo
  {journal} {Phys. Rev. B}\ }\textbf {\bibinfo {volume} {100}},\ \bibinfo
  {pages} {115124} (\bibinfo {year} {2019})}\BibitemShut {NoStop}%
\bibitem [{\citenamefont {Matsushita}\ \emph {et~al.}(2019)\citenamefont
  {Matsushita}, \citenamefont {Nagai},\ and\ \citenamefont
  {Fujimoto}}]{Matsushita_ER_PRB19}%
  \BibitemOpen
  \bibfield  {author} {\bibinfo {author} {\bibfnamefont {T.}~\bibnamefont
  {Matsushita}}, \bibinfo {author} {\bibfnamefont {Y.}~\bibnamefont {Nagai}}, \
  and\ \bibinfo {author} {\bibfnamefont {S.}~\bibnamefont {Fujimoto}},\ }\href
  {\doibase 10.1103/PhysRevB.100.245205} {\bibfield  {journal} {\bibinfo
  {journal} {Phys. Rev. B}\ }\textbf {\bibinfo {volume} {100}},\ \bibinfo
  {pages} {245205} (\bibinfo {year} {2019})}\BibitemShut {NoStop}%
\bibitem [{\citenamefont {Michishita}\ \emph {et~al.}(2020)\citenamefont
  {Michishita}, \citenamefont {Yoshida},\ and\ \citenamefont
  {Peters}}]{Michishita_EP_PRB20}%
  \BibitemOpen
  \bibfield  {author} {\bibinfo {author} {\bibfnamefont {Y.}~\bibnamefont
  {Michishita}}, \bibinfo {author} {\bibfnamefont {T.}~\bibnamefont {Yoshida}},
  \ and\ \bibinfo {author} {\bibfnamefont {R.}~\bibnamefont {Peters}},\ }\href
  {\doibase 10.1103/PhysRevB.101.085122} {\bibfield  {journal} {\bibinfo
  {journal} {Phys. Rev. B}\ }\textbf {\bibinfo {volume} {101}},\ \bibinfo
  {pages} {085122} (\bibinfo {year} {2020})}\BibitemShut {NoStop}%
\bibitem [{\citenamefont {Michishita}\ and\ \citenamefont
  {Peters}(2020)}]{Michishita_EP_PRL20}%
  \BibitemOpen
  \bibfield  {author} {\bibinfo {author} {\bibfnamefont {Y.}~\bibnamefont
  {Michishita}}\ and\ \bibinfo {author} {\bibfnamefont {R.}~\bibnamefont
  {Peters}},\ }\href {\doibase 10.1103/PhysRevLett.124.196401} {\bibfield
  {journal} {\bibinfo  {journal} {Phys. Rev. Lett.}\ }\textbf {\bibinfo
  {volume} {124}},\ \bibinfo {pages} {196401} (\bibinfo {year}
  {2020})}\BibitemShut {NoStop}%
\bibitem [{\citenamefont {Rausch}\ \emph {et~al.}(2020)\citenamefont {Rausch},
  \citenamefont {Peters},\ and\ \citenamefont {Yoshida}}]{Rausch_EP1D_arXiv20}%
  \BibitemOpen
  \bibfield  {author} {\bibinfo {author} {\bibfnamefont {R.}~\bibnamefont
  {Rausch}}, \bibinfo {author} {\bibfnamefont {R.}~\bibnamefont {Peters}}, \
  and\ \bibinfo {author} {\bibfnamefont {T.}~\bibnamefont {Yoshida}},\
  }\href@noop {} {\bibfield  {journal} {\bibinfo  {journal} {arXiv preprint
  arXiv:2008.03907}\ } (\bibinfo {year} {2020})}\BibitemShut {NoStop}%
\bibitem [{\citenamefont {Matsushita}\ \emph {et~al.}(2020)\citenamefont
  {Matsushita}, \citenamefont {Nagai},\ and\ \citenamefont
  {Fujimoto}}]{Matsushita_nHResp_arXiv20}%
  \BibitemOpen
  \bibfield  {author} {\bibinfo {author} {\bibfnamefont {T.}~\bibnamefont
  {Matsushita}}, \bibinfo {author} {\bibfnamefont {Y.}~\bibnamefont {Nagai}}, \
  and\ \bibinfo {author} {\bibfnamefont {S.}~\bibnamefont {Fujimoto}},\
  }\href@noop {} {\bibfield  {journal} {\bibinfo  {journal} {arXiv preprint
  arXiv:2004.11014}\ } (\bibinfo {year} {2020})}\BibitemShut {NoStop}%
\bibitem [{\citenamefont {Okuma}\ and\ \citenamefont
  {Sato}(2020{\natexlab{b}})}]{Okuma_PS_aXiv20208}%
  \BibitemOpen
  \bibfield  {author} {\bibinfo {author} {\bibfnamefont {N.}~\bibnamefont
  {Okuma}}\ and\ \bibinfo {author} {\bibfnamefont {M.}~\bibnamefont {Sato}},\
  }\href@noop {} {\bibfield  {journal} {\bibinfo  {journal} {arXiv preprint
  arXiv:2008.06498}\ } (\bibinfo {year} {2020}{\natexlab{b}})}\BibitemShut
  {NoStop}%
\bibitem [{\citenamefont {Metzner}\ and\ \citenamefont
  {Vollhardt}(1989)}]{WMetznerPRL89_DMFT}%
  \BibitemOpen
  \bibfield  {author} {\bibinfo {author} {\bibfnamefont {W.}~\bibnamefont
  {Metzner}}\ and\ \bibinfo {author} {\bibfnamefont {D.}~\bibnamefont
  {Vollhardt}},\ }\href {\doibase 10.1103/PhysRevLett.62.324} {\bibfield
  {journal} {\bibinfo  {journal} {Phys. Rev. Lett.}\ }\textbf {\bibinfo
  {volume} {62}},\ \bibinfo {pages} {324} (\bibinfo {year} {1989})}\BibitemShut
  {NoStop}%
\bibitem [{\citenamefont {M{\"u}ller-Hartmann}(1989)}]{MHartmannZP89_DMFT}%
  \BibitemOpen
  \bibfield  {author} {\bibinfo {author} {\bibfnamefont {E.}~\bibnamefont
  {M{\"u}ller-Hartmann}},\ }\href {\doibase 10.1007/BF01311397} {\bibfield
  {journal} {\bibinfo  {journal} {Zeitschrift f{\"u}r Physik B Condensed
  Matter}\ }\textbf {\bibinfo {volume} {74}},\ \bibinfo {pages} {507} (\bibinfo
  {year} {1989})}\BibitemShut {NoStop}%
\bibitem [{\citenamefont {Georges}\ \emph {et~al.}(1996)\citenamefont
  {Georges}, \citenamefont {Kotliar}, \citenamefont {Krauth},\ and\
  \citenamefont {Rozenberg}}]{AGeorgesRMP96_DMFT}%
  \BibitemOpen
  \bibfield  {author} {\bibinfo {author} {\bibfnamefont {A.}~\bibnamefont
  {Georges}}, \bibinfo {author} {\bibfnamefont {G.}~\bibnamefont {Kotliar}},
  \bibinfo {author} {\bibfnamefont {W.}~\bibnamefont {Krauth}}, \ and\ \bibinfo
  {author} {\bibfnamefont {M.~J.}\ \bibnamefont {Rozenberg}},\ }\href {\doibase
  10.1103/RevModPhys.68.13} {\bibfield  {journal} {\bibinfo  {journal} {Rev.
  Mod. Phys.}\ }\textbf {\bibinfo {volume} {68}},\ \bibinfo {pages} {13}
  (\bibinfo {year} {1996})}\BibitemShut {NoStop}%
\bibitem [{\citenamefont {Georges}\ and\ \citenamefont
  {Kotliar}(1992)}]{Georges_IPT_PRB92}%
  \BibitemOpen
  \bibfield  {author} {\bibinfo {author} {\bibfnamefont {A.}~\bibnamefont
  {Georges}}\ and\ \bibinfo {author} {\bibfnamefont {G.}~\bibnamefont
  {Kotliar}},\ }\href {\doibase 10.1103/PhysRevB.45.6479} {\bibfield  {journal}
  {\bibinfo  {journal} {Phys. Rev. B}\ }\textbf {\bibinfo {volume} {45}},\
  \bibinfo {pages} {6479} (\bibinfo {year} {1992})}\BibitemShut {NoStop}%
\bibitem [{\citenamefont {Zhang}\ \emph {et~al.}(1993)\citenamefont {Zhang},
  \citenamefont {Rozenberg},\ and\ \citenamefont {Kotliar}}]{Zhang_IPT_PRL93}%
  \BibitemOpen
  \bibfield  {author} {\bibinfo {author} {\bibfnamefont {X.~Y.}\ \bibnamefont
  {Zhang}}, \bibinfo {author} {\bibfnamefont {M.~J.}\ \bibnamefont
  {Rozenberg}}, \ and\ \bibinfo {author} {\bibfnamefont {G.}~\bibnamefont
  {Kotliar}},\ }\href {\doibase 10.1103/PhysRevLett.70.1666} {\bibfield
  {journal} {\bibinfo  {journal} {Phys. Rev. Lett.}\ }\textbf {\bibinfo
  {volume} {70}},\ \bibinfo {pages} {1666} (\bibinfo {year}
  {1993})}\BibitemShut {NoStop}%
\bibitem [{\citenamefont {Kajueter}\ and\ \citenamefont
  {Kotliar}(1996)}]{Kajueter_modIPT_96}%
  \BibitemOpen
  \bibfield  {author} {\bibinfo {author} {\bibfnamefont {H.}~\bibnamefont
  {Kajueter}}\ and\ \bibinfo {author} {\bibfnamefont {G.}~\bibnamefont
  {Kotliar}},\ }\href {\doibase 10.1103/PhysRevLett.77.131} {\bibfield
  {journal} {\bibinfo  {journal} {Phys. Rev. Lett.}\ }\textbf {\bibinfo
  {volume} {77}},\ \bibinfo {pages} {131} (\bibinfo {year} {1996})}\BibitemShut
  {NoStop}%
\bibitem [{\citenamefont {Trefethen}\ and\ \citenamefont
  {Embree}(2005)}]{Trefethen_psbook_2005}%
  \BibitemOpen
  \bibfield  {author} {\bibinfo {author} {\bibfnamefont {L.~N.}\ \bibnamefont
  {Trefethen}}\ and\ \bibinfo {author} {\bibfnamefont {M.}~\bibnamefont
  {Embree}},\ }\href@noop {} {\emph {\bibinfo {title} {Spectra and
  pseudospectra: the behavior of nonnormal matrices and operators}}}\ (\bibinfo
   {publisher} {Princeton University Press},\ \bibinfo {year}
  {2005})\BibitemShut {NoStop}%
\bibitem [{sub()}]{sublatsymm_ftnt}%
  \BibitemOpen
  \href@noop {} {}\bibinfo {note} {{ The symmetry constraint is called
  sublattice symmetry in Ref.~~\onlinecite{Kawabata_gapped_PRX19}. We, thus,
  follow this notation although our system does not have sublattice;
  $\alpha=a,b$ denotes orbital. }}\BibitemShut {NoStop}%
\bibitem [{\citenamefont {Schuricht}\ \emph {et~al.}(2012)\citenamefont
  {Schuricht}, \citenamefont {Andergassen},\ and\ \citenamefont
  {Meden}}]{Schurcht_LSW_JoP2012}%
  \BibitemOpen
  \bibfield  {author} {\bibinfo {author} {\bibfnamefont {D.}~\bibnamefont
  {Schuricht}}, \bibinfo {author} {\bibfnamefont {S.}~\bibnamefont
  {Andergassen}}, \ and\ \bibinfo {author} {\bibfnamefont {V.}~\bibnamefont
  {Meden}},\ }\href {\doibase 10.1088/0953-8984/25/1/014003} {\bibfield
  {journal} {\bibinfo  {journal} {Journal of Physics: Condensed Matter}\
  }\textbf {\bibinfo {volume} {25}},\ \bibinfo {pages} {014003} (\bibinfo
  {year} {2012})}\BibitemShut {NoStop}%
\bibitem [{LSW()}]{LSW_ftnt}%
  \BibitemOpen
  \href@noop {} {}\bibinfo {note} {{ The local spectral weight (, or the local
  density of states) is defined as~\cite{Schurcht_LSW_JoP2012} $ A(\omega)=
  \frac{1}{\sqrt{L_xL_y}}\sum_{\bm{k},\alpha}e^{i\bm{k}\cdot(\bm{R}-\bm{R}')}\frac{1}{\omega+i\delta
  -E_{\alpha}(\omega,\bm{k})} $ where $E_{\alpha}(\omega,\bm{k})$ is the
  eigenvalues of the effective Hamiltonian $\hat{H}(\omega,\bm{k})$. The
  vectors $\bm{R}$ and $\bm{R}'$ specifies the sites of the lattice.
  }}\BibitemShut {NoStop}%
\bibitem [{\citenamefont {Kotliar}\ \emph {et~al.}(2001)\citenamefont
  {Kotliar}, \citenamefont {Savrasov}, \citenamefont {P\'alsson},\ and\
  \citenamefont {Biroli}}]{Kotliar_CDMFT_PRL01}%
  \BibitemOpen
  \bibfield  {author} {\bibinfo {author} {\bibfnamefont {G.}~\bibnamefont
  {Kotliar}}, \bibinfo {author} {\bibfnamefont {S.~Y.}\ \bibnamefont
  {Savrasov}}, \bibinfo {author} {\bibfnamefont {G.}~\bibnamefont {P\'alsson}},
  \ and\ \bibinfo {author} {\bibfnamefont {G.}~\bibnamefont {Biroli}},\ }\href
  {\doibase 10.1103/PhysRevLett.87.186401} {\bibfield  {journal} {\bibinfo
  {journal} {Phys. Rev. Lett.}\ }\textbf {\bibinfo {volume} {87}},\ \bibinfo
  {pages} {186401} (\bibinfo {year} {2001})}\BibitemShut {NoStop}%
\bibitem [{\citenamefont {Maier}\ \emph {et~al.}(2005)\citenamefont {Maier},
  \citenamefont {Jarrell}, \citenamefont {Pruschke},\ and\ \citenamefont
  {Hettler}}]{Maier_CDMFT_RMP05}%
  \BibitemOpen
  \bibfield  {author} {\bibinfo {author} {\bibfnamefont {T.}~\bibnamefont
  {Maier}}, \bibinfo {author} {\bibfnamefont {M.}~\bibnamefont {Jarrell}},
  \bibinfo {author} {\bibfnamefont {T.}~\bibnamefont {Pruschke}}, \ and\
  \bibinfo {author} {\bibfnamefont {M.~H.}\ \bibnamefont {Hettler}},\ }\href
  {\doibase 10.1103/RevModPhys.77.1027} {\bibfield  {journal} {\bibinfo
  {journal} {Rev. Mod. Phys.}\ }\textbf {\bibinfo {volume} {77}},\ \bibinfo
  {pages} {1027} (\bibinfo {year} {2005})}\BibitemShut {NoStop}%
\bibitem [{\citenamefont {Wilson}(1975)}]{KWilsonRMP75_NRG}%
  \BibitemOpen
  \bibfield  {author} {\bibinfo {author} {\bibfnamefont {K.~G.}\ \bibnamefont
  {Wilson}},\ }\href {\doibase 10.1103/RevModPhys.47.773} {\bibfield  {journal}
  {\bibinfo  {journal} {Rev. Mod. Phys.}\ }\textbf {\bibinfo {volume} {47}},\
  \bibinfo {pages} {773} (\bibinfo {year} {1975})}\BibitemShut {NoStop}%
\bibitem [{\citenamefont {Peters}\ \emph {et~al.}(2006)\citenamefont {Peters},
  \citenamefont {Pruschke},\ and\ \citenamefont {Anders}}]{RPetersPRB06_NRG}%
  \BibitemOpen
  \bibfield  {author} {\bibinfo {author} {\bibfnamefont {R.}~\bibnamefont
  {Peters}}, \bibinfo {author} {\bibfnamefont {T.}~\bibnamefont {Pruschke}}, \
  and\ \bibinfo {author} {\bibfnamefont {F.~B.}\ \bibnamefont {Anders}},\
  }\href {\doibase 10.1103/PhysRevB.74.245114} {\bibfield  {journal} {\bibinfo
  {journal} {Phys. Rev. B}\ }\textbf {\bibinfo {volume} {74}},\ \bibinfo
  {pages} {245114} (\bibinfo {year} {2006})}\BibitemShut {NoStop}%
\bibitem [{\citenamefont {Bulla}\ \emph {et~al.}(2008)\citenamefont {Bulla},
  \citenamefont {Costi},\ and\ \citenamefont {Pruschke}}]{RBullaRMP08_NRG}%
  \BibitemOpen
  \bibfield  {author} {\bibinfo {author} {\bibfnamefont {R.}~\bibnamefont
  {Bulla}}, \bibinfo {author} {\bibfnamefont {T.~A.}\ \bibnamefont {Costi}}, \
  and\ \bibinfo {author} {\bibfnamefont {T.}~\bibnamefont {Pruschke}},\ }\href
  {\doibase 10.1103/RevModPhys.80.395} {\bibfield  {journal} {\bibinfo
  {journal} {Rev. Mod. Phys.}\ }\textbf {\bibinfo {volume} {80}},\ \bibinfo
  {pages} {395} (\bibinfo {year} {2008})}\BibitemShut {NoStop}%
\bibitem [{\citenamefont {Werner}\ \emph {et~al.}(2006)\citenamefont {Werner},
  \citenamefont {Comanac}, \citenamefont {de' Medici}, \citenamefont {Troyer},\
  and\ \citenamefont {Millis}}]{Werner_CTQMC_PRL06}%
  \BibitemOpen
  \bibfield  {author} {\bibinfo {author} {\bibfnamefont {P.}~\bibnamefont
  {Werner}}, \bibinfo {author} {\bibfnamefont {A.}~\bibnamefont {Comanac}},
  \bibinfo {author} {\bibfnamefont {L.}~\bibnamefont {de' Medici}}, \bibinfo
  {author} {\bibfnamefont {M.}~\bibnamefont {Troyer}}, \ and\ \bibinfo {author}
  {\bibfnamefont {A.~J.}\ \bibnamefont {Millis}},\ }\href {\doibase
  10.1103/PhysRevLett.97.076405} {\bibfield  {journal} {\bibinfo  {journal}
  {Phys. Rev. Lett.}\ }\textbf {\bibinfo {volume} {97}},\ \bibinfo {pages}
  {076405} (\bibinfo {year} {2006})}\BibitemShut {NoStop}%
\bibitem [{\citenamefont {Werner}\ and\ \citenamefont
  {Millis}(2006)}]{Werner_CTQMC_PRB06}%
  \BibitemOpen
  \bibfield  {author} {\bibinfo {author} {\bibfnamefont {P.}~\bibnamefont
  {Werner}}\ and\ \bibinfo {author} {\bibfnamefont {A.~J.}\ \bibnamefont
  {Millis}},\ }\href {\doibase 10.1103/PhysRevB.74.155107} {\bibfield
  {journal} {\bibinfo  {journal} {Phys. Rev. B}\ }\textbf {\bibinfo {volume}
  {74}},\ \bibinfo {pages} {155107} (\bibinfo {year} {2006})}\BibitemShut
  {NoStop}%
\bibitem [{\citenamefont {Haule}(2007)}]{Haule_CTQMC_PRB07}%
  \BibitemOpen
  \bibfield  {author} {\bibinfo {author} {\bibfnamefont {K.}~\bibnamefont
  {Haule}},\ }\href {\doibase 10.1103/PhysRevB.75.155113} {\bibfield  {journal}
  {\bibinfo  {journal} {Phys. Rev. B}\ }\textbf {\bibinfo {volume} {75}},\
  \bibinfo {pages} {155113} (\bibinfo {year} {2007})}\BibitemShut {NoStop}%
\end{thebibliography}
